\documentclass[fleqn,usenatbib]{mnras}

\usepackage{newtxtext,newtxmath}

\usepackage[T1]{fontenc}

\DeclareRobustCommand{\VAN}[3]{#2}
\let\VANthebibliography\thebibliography
\def\thebibliography{\DeclareRobustCommand{\VAN}[3]{##3}\VANthebibliography}

\usepackage{graphicx}	
\usepackage{amsmath}	
\usepackage{natbib}
\usepackage{threeparttable}

\newcommand{\Lya}{{\rm Ly}$\alpha$ }
\newcommand{\Ha}{{\rm H}$\alpha$}
\newcommand{\Hi}{H\,{\sc i}}
\newcommand{\fesca}{$f_{\mathrm{esc}}^{\mathrm{Ly\alpha}}$}
\newcommand{\fescc}{$f_{\mathrm{esc}}^{\mathrm{LyC}}$}
\newcommand{\fescmed}{$f_{\mathrm{esc,med}}^{\mathrm{Ly\alpha}}$}
\newcommand{\ewLya}{$\rm EW_0(Ly\alpha)$}

\newcommand{\nion}{$\dot{n}_{\mathrm{ion}}$}

\title[Ly$\alpha$ escape fraction at $z\simeq6.2$]{Subaru meets JWST: A Direct Measurement of Ly$\boldsymbol{\alpha}$ Escape Fraction at $\boldsymbol{z\simeq6.2}$ with Dual Narrow-Band Imaging}

\author[S. Shimizu et al.]{
Shunta Shimizu,$^{1}$\thanks{E-mail: s.shimizu@astron.s.u-tokyo.ac.jp}
Nobunari Kashikawa,$^{1,2}$
Ryo Emori,$^{1}$
Junya Arita,$^{1}$
Kohei Inayoshi,$^{3}$
\newauthor
Akio K. Inoue,$^{4,5}$
Kei Ito,$^{6,7}$
Satoshi Kikuta,$^{8}$
Kentaro Koretomo$^{1}$,
Mariko Kubo,$^{9,10}$,
\newauthor
Yongming Liang,$^{11,12}$,
Rieko Momose,$^{11,13,14}$
Kentaro Nagamine,$^{14,15,16,17,18}$
Masafusa Onoue,$^{14,19}$
\newauthor
Rhythm Shimakawa,$^{19}$
Yoshihiro Takeda,$^{1}$
Hisakazu Uchiyama,$^{11,20}$
and Takehiro Yoshioka$^{1}$
\\
$^{1}$Department of Astronomy, School of Science, The University of Tokyo, 7-3-1 Hongo, Bunkyo, Tokyo 113-0033, Japan\\
$^{2}$Research Center for the Early Universe, The University of Tokyo, 7-3-1 Hongo, Bunkyo, Tokyo 113-0033, Japan\\
$^{3}$Kavli Institute for Astronomy and Astrophysics, Peking University, Beijing 100871, China\\
$^{4}$Waseda Research Institute for Science and Engineering, Faculty of Science and Engineering, Waseda University, 3-4-1 Okubo, Shinjuku, \\Tokyo 169-8555, Japan\\
$^{5}$Department of Physics, School of Advanced Science and Engineering, Faculty of Science and Engineering, Waseda University, 3-4-1 Okubo,\\ Shinjuku, Tokyo 169-8555, Japan\\
$^{6}$Cosmic Dawn Center (DAWN), Copenhagen, Denmark \\
$^{7}$DTU Space, Technical University of Denmark, Elektrovej 327, DK2800 Kgs. Lyngby, Denmark\\
$^{8}$Department of Regional Promotion, Nara Prefectural University, 10 Funahashicho, Nara, Nara 630-8258, Japan\\
$^{9}$Astronomical Institute, Tohoku University, 6-3, Aramaki, Aoba, Sendai, Miyagi 980-8578, Japan \\
$^{10}$Department of Physics and Astronomy, School of Science, Kwansei Gakuin University, 1 Gakuen Uegahara, Sanda, Hyogo 669-1330, Japan\\
$^{11}$National Astronomical Observatory of Japan, Mitaka, Tokyo 181-8588, Japan\\
$^{12}$Institute for Cosmic Ray Research, The University of Tokyo, 5-1-5 Kashiwanoha, Kashiwa, Chiba 277-8582, Japan\\
$^{13}$Carnegie Observatories, 813 Santa Barbara Street, Pasadena, CA 91101, USA\\
$^{14}$Kavli Institute for the Physics and Mathematics of the Universe (Kavli IPMU, WPI), UTIAS, The University of Tokyo, 5-1-5 Kashiwanoha,\\ Kashiwa, Chiba 277-8583, Japan\\
$^{15}$Theoretical Astrophysics, Department of Earth and Space Science, Graduate School of Science, The University of Osaka, 1-1 Machikaneyama,\\ Toyonaka, Osaka 560-0043, Japan\\
$^{16}$Theoretical Joint Research, Forefront Research Center, Graduate School of Science, The University of Osaka, Toyonaka, Osaka 560-0043, Japan\\
$^{17}$Department of Physics and Astronomy, University of Nevada, Las Vegas, 4505 S. Maryland Pkwy, Las Vegas, NV 89154-4002, USA\\
$^{18}$Nevada Center for Astrophysics, University of Nevada, Las Vegas, 4505 S. Maryland Pkwy, Las Vegas, NV 89154-4002, USA\\
$^{19}$Waseda Institute for Advanced Study (WIAS), Waseda University, 1-21-1, Nishi-Waseda, Shinjuku, Tokyo 169-0051, Japan\\
$^{20}$Department of Advanced Sciences, Faculty of Science and Engineering, Hosei University, 3-7-2 Kajino-cho, Koganei, Tokyo 184-8584, Japan
}

\date{Accepted XXX. Received YYY; in original form ZZZ}

\pubyear{\the\year{}}

\begin{document}
\label{firstpage}
\pagerange{\pageref{firstpage}--\pageref{lastpage}}
\maketitle

\begin{abstract}
We present a direct measurement of the Ly$\alpha$ escape fraction, $f^{\rm Ly\alpha}_{\rm esc}$, for H$\alpha$ emitters (HAEs) at $z\simeq6.2$ in the JWST CEERS field by combining JWST/NIRCam F470N imaging with Subaru/HSC NB872 imaging.
This unique pair of narrow-band filters enables the simultaneous measurement of H$\alpha$ and Ly$\alpha$ fluxes from galaxies during the epoch of reionization (EoR).
We select 84 HAEs from F470N excesses, among which 56 have reliable NB872 photometry and 14 are detected in Ly$\alpha$.
The completeness-weighted stack of the HAE sample yields a median $f^{\rm Ly\alpha}_{\rm esc}$ at $z\simeq6.2$ of $0.106^{+0.066}_{-0.044}$, which is in good agreement with recent measurements at similar redshifts.
We further find that the stacked $f_{\rm esc}^{\rm Ly\alpha}$ shows no statistically significant change as the H$\alpha$ luminosity threshold is lowered over the range probed by our sample.
If Ly$\alpha$ escape traces Lyman continuum leakage,
this may suggest that 
relatively H$\alpha$-luminous HAEs, 
rather than only the faintest ones, can provide an important contribution to the ionizing photon budget during the EoR.
For individual galaxies, $f^{\rm Ly\alpha}_{\rm esc}$
positively correlates with Ly$\alpha$ equivalent width and negatively correlates with the UV continuum slope $\beta$.
We also find a tentative trend toward higher
$f^{\rm Ly\alpha}_{\rm esc}$ at smaller rest-frame UV
sizes, but no significant correlation with SED-derived
$E(B-V)$ or rest-frame optical size.
Although based on a limited sample, these results suggest that the galaxy-to-galaxy variation in $f_{\rm esc}^{\rm Ly\alpha}$ 
may be more closely
linked to 
compact, low-attenuation star-forming components traced by the UV continuum than to global dust attenuation or the overall stellar structure.
\end{abstract}




\graphicspath{{./fig_shimizu26/}}

\section{Introduction}
\Lya is the resonant transition of neutral hydrogen (\Hi) at a rest-frame wavelength of 1215.67\,\AA, corresponding to the $2p \rightarrow 1s$ transition, and is one of the most prominent emission lines in the Universe.
Following the early prediction of strong \Lya emission from young galaxies \citep{Partridge+67}, \Lya has long played a central role in the discovery and confirmation of high-$z$ galaxies, as well as in studies of their physical properties and the progress of cosmic reionization (e.g. \citealt{Hu+98,Gawiser+06,Kashikawa+06,Ouchi+08,Ouchi+10}).
The particular importance of \Lya stems from its resonant nature.
After being produced in $\mathrm{H\,\text{\textsc{ii}}}$ regions, \Lya photons undergo repeated resonant scattering by \Hi\ within and around galaxies, and at sufficiently high redshift, they can be further attenuated by the partially neutral intergalactic medium (IGM).
This process greatly increases their effective path lengths, making them highly susceptible to even small amounts of dust.
In addition, the \Hi\ covering fraction, gas kinematics, geometry, and viewing angle can strongly affect the \Lya line flux, spectral profile, and spatial distribution (e.g. \citealt{Verhamme+06, Dijkstra+14, Herenz+25, AlmadaMonter+26}).
Consequently, the observed \Lya flux cannot in general be translated directly into the star-formation rate (SFR) or ionizing photon production rate (e.g. \citealt{Atek+08, Hayes+10}).
Moreover, Ly$\alpha$-selected galaxies do not represent an unbiased subset of the parent star-forming population, but are often biased toward blue, low-mass, and relatively dust-poor systems (e.g. \citealt{Gawiser+07, Ono+10b,Haro+20,Iani+24}).

A commonly used anchor for interpreting these complex radiative-transfer effects is the non-resonant hydrogen recombination line \Ha.
Because \Ha\ directly traces nebular recombination in ionized gas and is not subject to strong resonant scattering like Ly$\alpha$, it provides a useful estimate of the intrinsic \Lya photon production rate
associated with star formation.
Under Case B recombination, the intrinsic Ly$\alpha$/H$\alpha$ ratio is commonly taken to be 8.7 for typical nebular conditions \citep{Hummer}; deviations from this value reflect the effects of \Hi\ scattering, dust absorption, gas kinematics, geometry, and viewing angle (e.g. \citealt{Laursen+09,Hayes+15}).
Direct comparisons between \Ha\ and \Lya in the same galaxies therefore provide a powerful way to quantify the \Lya escape fraction, \fesca, and to investigate its physical origin (e.g. \citealt{Oteo+15, Matthee+16,Shimakawa+17,Gazagnes+20,Lin+24, Shimizu+25}).
\citet{Oteo+15} combined \Ha-selected galaxies with existing \Lya data and 
found that only $\simeq 4.5\%$ of \Ha\ emitters (HAEs) show detectable \Lya emission at $z \sim 2$.
They also showed that \fesca\ decreases with increasing dust attenuation, 
redder UV continuum slope, stellar mass ($M_* $), and SFR. 
This indicates that \Lya escape  
is not governed by star-formation activity alone, but is strongly regulated by the dust content and the structure of the interstellar medium (ISM) and circumgalactic medium (CGM).

Such direct comparisons become increasingly difficult from the ground once \Ha\ is redshifted beyond the $K$ band at $z \gtrsim 2.5$.
The \textit{James Webb Space Telescope} (\textit{JWST}, \citealt{JWST}) has dramatically transformed this situation.
In particular, NIRCam wide-field slitless spectroscopy (WFSS) covers 2.4--5.0 $\mu$m and NIRSpec covers 0.6--5.3 $\mu$m, enabling direct access to \Ha\ at much higher redshift, even during the epoch of reionization (EoR).
Using the NIRCam/WFSS data together with deep \Lya spectroscopy, \citet{Tang+24a} characterized the \fesca\ distributions of galaxies at $z\simeq5$--6.
They found that galaxies with large \fesca\ are relatively common at these redshifts, and that \Lya escape tends to be more efficient in UV-fainter, bluer galaxies with stronger rest-frame optical
emission. 
These results suggest that \Lya escape at the end of
reionization is closely connected to young, low-dust, intense star-forming systems.
Recent studies based on \textit{JWST} observations have also reported that \fesca\ tends to be higher in UV-fainter, bluer galaxies (e.g. \citealt{Ning+23,Ning+26,Lin+24,Saxena+24}). 

In addition to object-by-object measurements, the cosmic-averaged \Lya escape fraction is a key quantity for interpreting \Lya observations during the EoR.
A conventional approach to estimating the cosmic-averaged \fesca\ involves comparing the Ly$\alpha$ luminosity function (LF) with the UV LF (e.g. \citealt{Hayes+11,Konno+16,Goovaerts+24b}). 
With the advent of \textit{JWST}, it is now feasible to measure the cosmic-averaged \fesca\ even at $z>2$ by comparing the Ly$\alpha$ LF with the H$\alpha$ LF, which shares a similar physical timescale with Ly$\alpha$ emission \citep{Sun+23}.
However, LF-based methods suffer from systematic uncertainties, including dependencies on the assumed functional form of the LF and the choice of integration limits. 
Furthermore, it remains uncertain whether comparisons based on fixed multiples of the characteristic luminosity ($L^*$) across two different LFs truly trace the same underlying galaxy population.
These uncertainties motivate a direct measurement of the cosmic-averaged \fesca\ from the same \Ha-selected galaxies.
\citet{Lin+24} derived the median \fesca\ (\fescmed) at $z=4.9$--$6.3$ by stacking the \Ha\ and \Lya emission directly measured from NIRCam/WFSS data and VLT/MUSE observations, respectively. 
They obtained $f_{\rm esc, med}^{\rm Ly\alpha}=0.090\pm0.006$, providing a key spectroscopic benchmark for the cosmic-averaged \fesca\ at $z\simeq5$--6.

Narrow-band (NB) imaging for \Ha\ provides a complementary route to such measurements.
Because it is imaging-based, NB selection can efficiently identify line emitters within a well-defined wavelength interval while preserving two-dimensional spatial information. 
Unlike slit-based spectroscopy, NB imaging avoids slit losses and uncertainties associated with slit placement (e.g. \citealt{Jiang+24,Napolitano+24}), as well as the spectral overlap and extraction uncertainties encountered in slitless spectroscopy (e.g. \citealt{Covelo-Paz+25}).
Recent \textit{JWST}/NIRCam NB programmes demonstrate that, in specific wavelength windows,  the detection limits for line fluxes can be comparable to, or fainter than, those reached by existing slitless spectroscopic surveys \citep{Duncan+25}.
\textit{JWST} NIRCam is equipped with several NB filters, the longest of which is F470N (central wavelength $\lambda_c = 4.708\,\mathrm{\mu m}$, bandwidth $= 0.051\,\mathrm{\mu m}$), which can capture \Ha\ emission from galaxies at $z\simeq6.2$.

In this study, we show the first measurement of \fesca\ at $z\simeq6.2$ based on dual-NB imaging, using the NB872 filter (central wavelength $\lambda_c = 0.872\,\mathrm{\mu m}$, $\mathrm{bandwidth}= 0.0067\,\mathrm{\mu m}$) of Subaru/Hyper Suprime-Cam (HSC; \citealp{Miyazaki2018, Komiyama2018, Kawanomoto2018, Furusawa2018}) to capture the \Lya\ emission from the HAEs detected in F470N.
\begin{figure}
    \centering
    \includegraphics[width=\columnwidth]{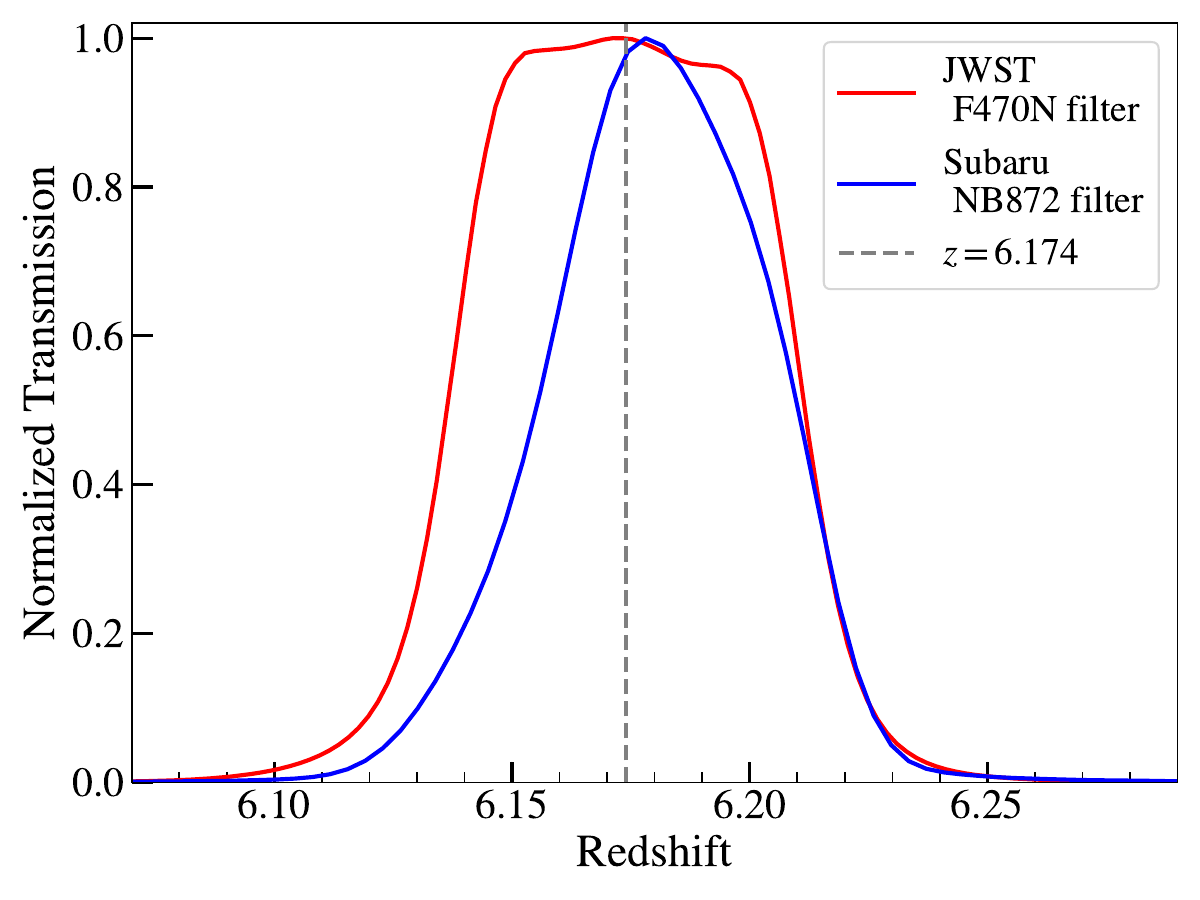}
    \caption{Normalized transmission curves of \textit{JWST}/NIRCam F470N (red) and Subaru/HSC NB872 (blue) filters. The horizontal axis is shown in terms of the redshift at which the \Lya emission line falls within NB872 and the \Ha\ emission line falls within F470N. The vertical dashed line indicates $z=6.174$, corresponding to the wavelength at the peak transmission of F470N.}
    \label{fig:filter_profiles}
\end{figure}
As shown in Figure~\ref{fig:filter_profiles}, the combination of JWST/NIRCam F470N and Subaru/HSC NB872 provides the only available dual-NB filter pair that can simultaneously capture \Ha\ and \Lya fluxes from the same galaxies at $z>6$, offering a rare observational window to directly measure both lines through NB imaging.
The redshift intervals corresponding to transmission above 50\% of the peak are $6.136<z<6.212$ for \Ha\ in F470N and $6.155<z<6.210$ for \Lya in NB872.
At lower redshift, \citet{Matthee+16} implemented a closely analogous
dual-NB approach at $z=2.23$, measuring \Lya emission for H$\alpha$-selected galaxies.
They derived both individual and median $f_{\rm esc}^{\rm Ly\alpha}$ and investigated their dependence on galaxy properties.
In this study, we apply the same technique to the EoR for the first time.

The structure of this paper is as follows. 
Section~\ref{sec:data} describes the imaging data used in this study, including the \textit{JWST}/NIRCam F470N, Subaru/HSC NB872, and ancillary \textit{Hubble Space Telescope (HST)} and \textit{JWST} observations, and presents the construction of the HAE sample at $z\simeq6.2$. 
Section~\ref{sec:method} describes the analysis methods, including the spectral energy distribution (SED) fitting, the measurement of \fesca\ from the Ly$\alpha$ and H$\alpha$ fluxes, the measurement of galaxy sizes, and the stacking analysis. 
In Section~\ref{sec:results_and_discussion}, we present the results and discussion. 
We first examine the physical properties of the HAEs in Section~\ref{sec:HAE_physical_properties}, and then present the \fescmed\ derived from stacking analysis and discuss its dependence on the adopted lower \Ha\ luminosity limit in Section~\ref{sec:stacking_HAE}. 
We then investigate the dependence of the individual \fesca\ on galaxy properties in Section~\ref{sec:corr_fesca}. 
Finally, we summarize our main conclusions in Section~\ref{sec:summary}.
Throughout this paper, we assume a flat $\Lambda$CDM cosmology with $H_0=70\,\mathrm{km\,s^{-1}\,Mpc^{-1}}$, $\Omega_\mathrm{m}=0.3$, and $\Omega_\Lambda=0.7$. 
All magnitudes in this paper refer to AB magnitude \citep{Oke}.

\section{Data} \label{sec:data}
\subsection{\textit{JWST}/NIRCam F470N imaging} \label{sec:F470N_data}
The Cosmic Evolution Early Release Science (CEERS) survey field \citep{Finkelstein+23} has been observed with \textit{JWST} using the F470N filter to detect HAEs at $z\simeq6.2$.
The F470N imaging is obtained as part of the Cycle 1 GO programme 2234 \citep{CEERS_F470N}.
We retrieve the uncalibrated (\texttt{uncal}) exposures from the Mikulski Archive for Space Telescopes (MAST) and reprocess them with the \textit{JWST} Science Calibration Pipeline \texttt{v1.16.1}, using the \texttt{jwst\_1303.pmap} Calibration Reference Data System (CRDS) file.
During the resample step, we set the output pixel scale to
$0^{\prime\prime}\!\!.03$ to match the pixel scale of the other \textit{HST} and \textit{JWST} broad-band data used in
Section~\ref{sec:other_data}.
In addition to the standard pipeline processing, we remove residual $1/f$ noise and subtract the sky background following the procedure adopted in \citet{Finkelstein+25} \footnote{\url{https://github.com/ceers}}.
We then refine the astrometric calibration and align the images to an external reference catalogue. 
To ensure that a sufficient number of reference sources are available within each NIRCam pointing, we use the same \textit{HST} based reference catalogue employed by the CEERS NIRCam imaging reduction \citep{Bagley+23}. 
This reference catalogue is derived from the \textit{HST}/WFC3 F160W mosaic in the Extended Groth Strip (EGS) field \citep{Grogin+11,Koekemoer+11}.
The final F470N mosaic covers a total area of 96.4 arcmin$^2$ in the CEERS field.
The point spread function (PSF) models are constructed by selecting $\sim 100$ isolated star-like sources, following the criteria outlined in \citet{Shimizu+25}, and stacking them using the \texttt{ePSF} module in \texttt{photutils} v2.0.2 \citep{photutils}.
The full width at half maximum (FWHM) of the PSF is measured from these empirical PSF models.
The 5$\sigma$ limiting magnitude is measured by masking bright sources and then calculating the standard deviation of the flux distribution 
obtained from randomly placed apertures in the background regions.
As shown in Table~\ref{tab:HAE_filter}, the FWHM of the PSF is 
$0^{\prime\prime}\!\!.175$. 
The $5\sigma$ limiting magnitudes measured in 
$0^{\prime\prime}\!\!.3$- and $0^{\prime\prime}\!\!.6$-diameter apertures are 
26.41 and 25.65, respectively.

\begin{table}
\centering
\caption{The effective wavelength ($\lambda_\mathrm{eff}$), $5\sigma$ depths, and FWHMs of the PSF for each filter used in this paper. For Subaru/HSC, the $5\sigma$ depth measurements are made using a $1^{\prime\prime}\!\!.5$-diameter aperture, and for \textit{HST} and \textit{JWST}, a $0^{\prime\prime}\!\!.6$-diameter aperture is used. }
\begin{threeparttable}
\begin{tabular}{llccc}
\hline
Instrument & Filter & $\lambda_\mathrm{eff}$ [$\mu$m] & 5$\sigma$ depth & FWHM [$\arcsec$] \\ \hline
\textit{HST}/ACS & F435W     & 0.434 & 27.62 & 0.100 \\
 & F606W  & 0.581  & 27.71  & 0.098\\
 & F814W  & 0.797  & 27.54  & 0.098\\ \hline
 Subaru/HSC & \textbf{NB872} & 0.873 & 25.87\tnote{$\dagger$} & 0.583 \\ \hline
\textit{JWST}/NIRCam & F090W &  0.899& 27.90 & 0.086\\
 & F115W & 1.143 & 28.05 & 0.065\\
 & F150W & 1.487 & 27.92 & 0.074\\
 & F200W & 1.986 & 28.13 & 0.082\\
 & F277W & 2.728 & 28.45 & 0.113\\
 & F356W & 3.529 & 28.53 & 0.139\\
 & F410M & 4.072 & 27.76 & 0.158\\
 & F444W & 4.350 & 28.17 & 0.159\\
 & \textbf{F470N} & 4.708 & 25.65\tnote{$\dagger$} & 0.175 \\ \hline
\end{tabular}
\begin{tablenotes}[flushleft]
\footnotesize
\item[$\dagger$] The corresponding $5\sigma$ line-flux limits are
$4.56\times10^{-18}$ and $1.38\times10^{-18}$
$\mathrm{erg\,s^{-1}\,cm^{-2}}$ for NB872 (Ly$\alpha$) and F470N (\Ha),
respectively, assuming negligible continuum emission and
emission lines that are much narrower than the filter bandwidths and centred at their peak transmission wavelengths.
\end{tablenotes}
\end{threeparttable}
\label{tab:HAE_filter}
\end{table}

\subsection{Subaru/HSC NB872 imaging} \label{sec:NB872_data}
We use the NB872 filter of Subaru/HSC to detect the Ly$\alpha$ emission from galaxies whose H$\alpha$ emission falls within the JWST/F470N at $z\simeq6.2$, as shown in Figure~\ref{fig:filter_profiles}.

The observations are conducted in the HSC queue mode in August 2023, from December 2024 to January 2025, and from March to April 2025 as shown in Table~\ref{tab:obs_info}.
\begin{table*} 
\centering
\caption{Observation information for Subaru/HSC NB872 imaging. }
\label{tab:obs_info}
\begin{threeparttable}
\begin{tabular}{lllccl}
\hline
Field & R.A. & Decl. & Exposure time [hr]\tnote{a} & Seeing [\arcsec]\tnote{b} & Observation date \\ 
\hline
CEERS & 14:19:18.00 & +52:49:30.0 & 1.75/1.50 & 1.35 (0.97--1.71) & 2023 Aug. \\
& & & 7.00/6.25 & 0.87 (0.56--1.34) & 2024 Dec. -- 2025 Jan.  \\
& & & 12.81/11.75 & 0.83 (0.50--1.72) & 2025 Mar. -- Apr. \\
\hline
\end{tabular}
\begin{tablenotes}
\footnotesize
\item[a] Exposure time denotes the total observing time / effective exposure time used for the analysis.
\item[b] Seeing is given as the mean value, with the range in parentheses.
\end{tablenotes} 
\end{threeparttable}
\end{table*}
The total observing time is 21.56 hours, but after excluding exposures affected by guide star tracking issues and ghost contamination, the effective exposure time used for the analysis is 19.5 hours (900\,s $\times$ 78\,shots).
Nine of these shots used in the analysis are taken under poor seeing conditions ($>1^{\prime\prime}\!\!.3$), but excluding them does not affect the final image quality.
The CCD021 malfunctioned on April 27, 2025; therefore, the data acquired from this CCD on and after this date are excluded from the data reduction.
The data are reduced using HSC pipeline (\texttt{hscPipe} \texttt{v8.5.3}) \citep{hscpipe}.
In this process, the zero point is automatically determined by matching stars in the field to the Pan-STARRS1 reference catalogue \citep{Panstarrs}. 
However, since this filter is newly used, we conduct a separate verification of the validity of this correction \citep{Arita+26}.
As a result, the photometric zero point is 27.015 mag, which differs from the value determined by \texttt{hscPipe} by only $\sim0.02$ mag (corresponding to $\sim2\%$ in flux), which should be negligible.
Although HSC has a very wide field of view ($\sim1.0\,\mathrm{deg}^2$), this study uses only the CEERS field imaged by F470N, which corresponds to about 10\% in the central area.
Within the area, the average FWHM of the PSF is $0^{\prime\prime}\!\!.58$, and the average 5$\sigma$ limiting magnitude is 25.87 in a $1^{\prime\prime}\!\!.5$-diameter aperture.

\subsection{Ancillary \textit{HST} and \textit{JWST} imaging} \label{sec:other_data}
In the CEERS field, multi-wavelength imaging surveys have been carried out with \textit{HST} and \textit{JWST} \citep{Grogin+11,Koekemoer+11,Finkelstein+25,Wang+25}. 
We use these imaging data to measure the continuum emission of HAEs at $z\simeq6.2$. 
Specifically, we employ 11 filters in total: \textit{HST}/F435W, F606W, and F814W, and \textit{JWST}/F090W, F115W, F150W, F200W, F277W, F356W, F410M, and F444W.
Among these bands, we use the reduced images from CEERS Data Release 1.0 (DR1), produced by \citet{Bagley+23} and \citet{Finkelstein+25}, except for F090W.
The F090W imaging data are obtained in the same GO programme as F470N, and reduced using the same procedure described in Section~\ref{sec:F470N_data}.
The PSF FWHMs and $5\sigma$ limiting magnitudes for each filter are summarized in Table~\ref{tab:HAE_filter}.

\subsection{Construction of the HAE sample}
\subsubsection{Multi-wavelength photometric catalogue} \label{sec:multi_catalog}
We construct a photometric catalogue using \textit{HST} and \textit{JWST} images homogenized to a common PSF, which facilitates the robust selection of emission line galaxies with NB filters.
For source detection, we adopt the F470N image as the detection band and perform forced photometry in the other bands after convolving each image to match the F470N PSF.
This approach is optimal since all the \textit{HST}/\textit{JWST} images used in this study have smaller PSF sizes than F470N (FWHM$\sim0^{\prime\prime}\!\!.175$).
Note that the HSC/NB872 data require a different photometric approach due to their much coarser spatial resolution compared to \textit{JWST}, which is described in detail in Section~\ref{sec:NB872_photometry}.
First, we construct PSF models for all \textit{HST} and \textit{JWST} images.
We then generate convolution kernels to match the PSF of each band to that of F470N using \textsc{pypher} \citep{pypher}. 
The convolution is performed using the \mbox{\textsc{photutils} \texttt{v2.0.2}} \citep{photutils}.
To validate the PSF homogenization, we compare the curves of growth of the PSF-matched images with that of F470N. 
The enclosed-flux fractions agree to within 2\% over the full range of aperture radii tested for all bands, indicating a sufficiently good PSF match for typical photometric applications (e.g. \citealt{Pirie+25}).
Next, we perform source detection on the F470N image using \textsc{SExtractor} \citep{sextractor}. 
We adopt detection parameters of \texttt{DETECT\_MINAREA}$=9$ and \texttt{DETECT\_THRESH}$=1.5$ and employ a Gaussian filter with an FWHM of 4 pixels ($0^{\prime\prime}\!\!.12$).
These parameters are determined after verification with small cutouts so that spectroscopically confirmed galaxies at $z\simeq6.2$ (see Section~\ref{sec:select_HAE}) can be detected in F470N, and spurious detections can be minimized. 

Finally, we perform forced photometry across all bands using \textsc{SExtractor} in dual-image mode, utilizing the F470N image as a detection band.
Following \citet{Pirie+25}, we adopt two aperture diameters for photometry, $0^{\prime\prime}\!\!.3$ and $0^{\prime\prime}\!\!.6$. 
We utilize the $0^{\prime\prime}\!\!.3$-diameter apertures for the selection of HAEs to maximize the signal-to-noise ratio (S/N), while we use the $0^{\prime\prime}\!\!.6$-diameter apertures for SED fitting to capture a larger fraction of the flux (approximating the total flux).
Photometric uncertainties estimated by propagating the \textit{JWST} pipeline error maps within an aperture may be systematically underestimated because pixel-to-pixel correlations in the resampling process are not fully accounted for in the error propagation (e.g. \citealt{Bagley+23}).
To address this issue, we perform an empirical noise analysis by placing 100,000 apertures at random positions in each band.
We measure the standard deviation of the fluxes in these apertures and compare it with the uncertainties derived from the error maps.
We find that the pipeline error maps underestimate the noise 
by a factor of $\sim3.5$ for F470N and $\sim1.5$ for the other CEERS bands.
The final uncertainties are obtained by rescaling the initial aperture uncertainties derived from the error maps by these correction factors, allowing them to retain the spatial variation of the local noise across the mosaics while accounting for the correlated-noise correction.

\subsubsection{Selection of NB excess sources}
We identify HAEs from the multi-wavelength catalogue by selecting sources showing a significant flux excess in the F470N relative to the broad-band continuum.
Prior to the selection, we apply several quality cuts to clean the catalogue. 
First, we exclude sources affected by saturation and those lying outside the valid imaging footprint in F444W and F470N.
To ensure a homogeneous selection, we also exclude sources in low-S/N regions near the footprint edges (25 pixels) in the F470N and F444W images. 
After applying these footprint and edge masks, the effective F470N-covered area used for the HAE selection is 85.2 arcmin$^2$.
We further exclude sources with S/N $<2$ in all bands other than F470N as likely spurious detections.
Applying these criteria yields a sample of 19,836 sources.

\begin{figure}
    \centering
    \includegraphics[width=\columnwidth]{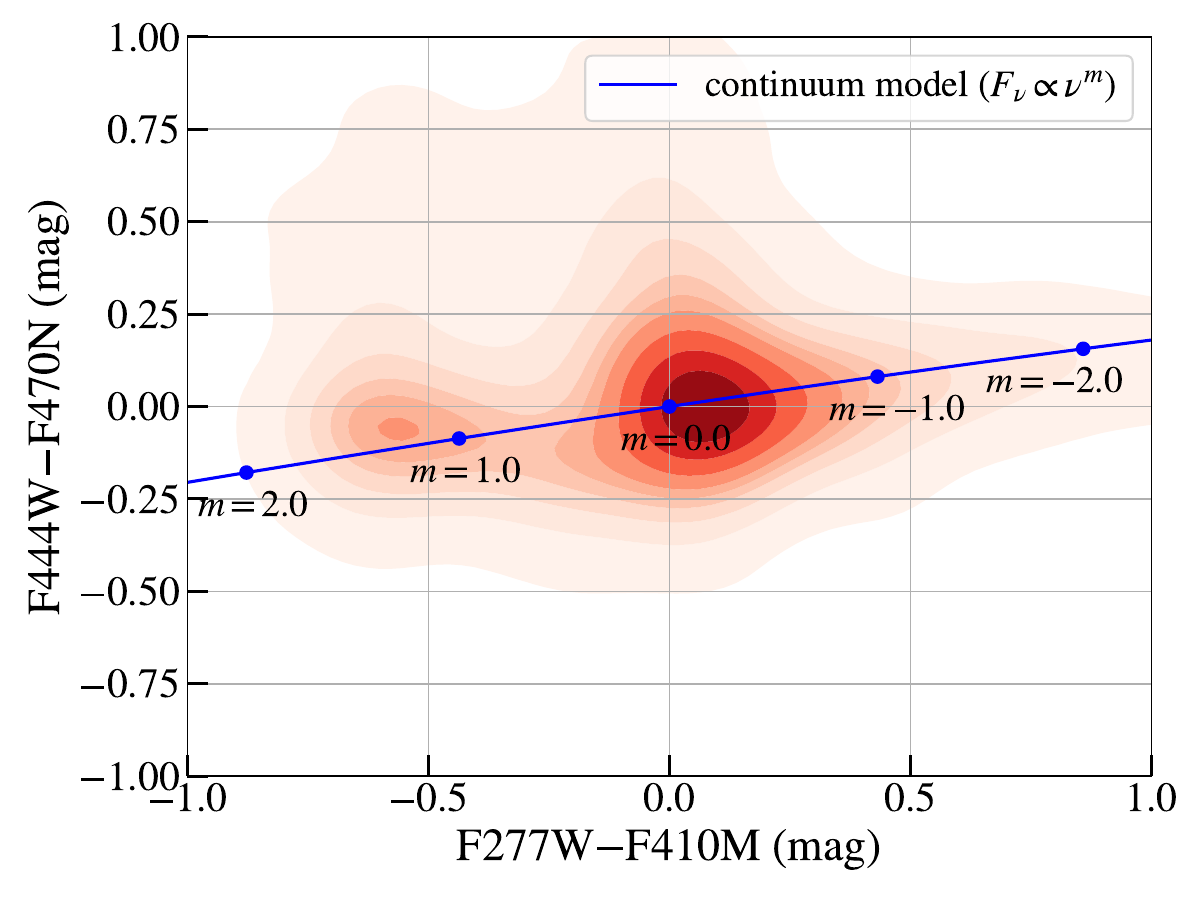}
    \caption{Relation between the observed continuum colour, $\mathrm{F277W}-\mathrm{F410M}$, and the expected continuum colour, $\mathrm{F444W}-\mathrm{F470N}$, for a power law spectrum $F_\nu \propto \nu^{m}$. The blue curve shows the model relation obtained by integrating power law spectra through the NIRCam filter transmission curves, with blue circles marking representative slopes $m=2, 1, 0, -1,$ and $-2$. The red contours show the distribution of observed colours for sources in the multi-wavelength catalogue constructed in Section~\ref{sec:multi_catalog}.}
    \label{fig:color_correction}
\end{figure}

When selecting NB excess sources, a non-flat continuum can bias the inferred NB excess because the NB and broad-band filters have different central wavelengths. 
We therefore estimate the continuum slope using two nearby filters and correct the observed ${\rm F444W}-{\rm F470N}$ colour (e.g. \citealt{Sobral+13, Pirie+25}). 
Although F444W is the broad-band closest to ${\rm F470N}$, it may be contaminated by H$\alpha$ emission for galaxies at $z\simeq6.2$ and thus cannot be used to measure the continuum slope. 
Likewise, F356W may be affected by [O\,{\sc iii}]$\lambda5007$ and is also unsuitable. 
We instead use F277W and F410M to estimate the continuum slope and correct ${\rm F444W}-{\rm F470N}$.
At $z\simeq6.2$, both filters probe the continuum redward of the Balmer break in terms of their effective wavelengths, and are not expected to be dominated by strong rest-frame optical emission lines.
We therefore approximate the continuum as a power law, $F_\nu \propto \nu^{m}$, and predict the expected continuum-only colour between F444W and F470N as 
\begin{equation}
    ({\rm F444W}-{\rm F470N})_{\rm cont}=f\!\left(({\rm F277W}-{\rm F410M})_{\rm obs}\right),
\end{equation}
where the function $f$ is constructed by integrating the model spectra through the filter transmission curves and linearly interpolating as shown in Figure~\ref{fig:color_correction}. 
This relation is broadly consistent with the distribution of sources in our catalogue, shown by the red contours in Figure~\ref{fig:color_correction}. 
The continuum-corrected colour, $({\rm F444W}-{\rm F470N})_{\rm corr}$ is then calculated as:
\begin{equation}
\begin{split}
({\rm F444W}-{\rm F470N})_{\rm corr}
&= ({\rm F444W}-{\rm F470N})_{\rm obs} \\
&\quad - f\!\left(({\rm F277W}-{\rm F410M})_{\rm obs}\right).
\end{split}
\end{equation}
For sources with S/N $< 2$ in either F277W or F410M, no correction is applied because the continuum slope can not be reliably estimated.
For sources with S/N $< 2$ in F444W, we adopt the $2\sigma$ limiting flux as an upper limit and use the corresponding lower limit on $(\mathrm{F444W}-\mathrm{F470N})_{\rm corr}$ for the F470N excess selection.

\begin{figure}
    \centering
    \includegraphics[width=\columnwidth]{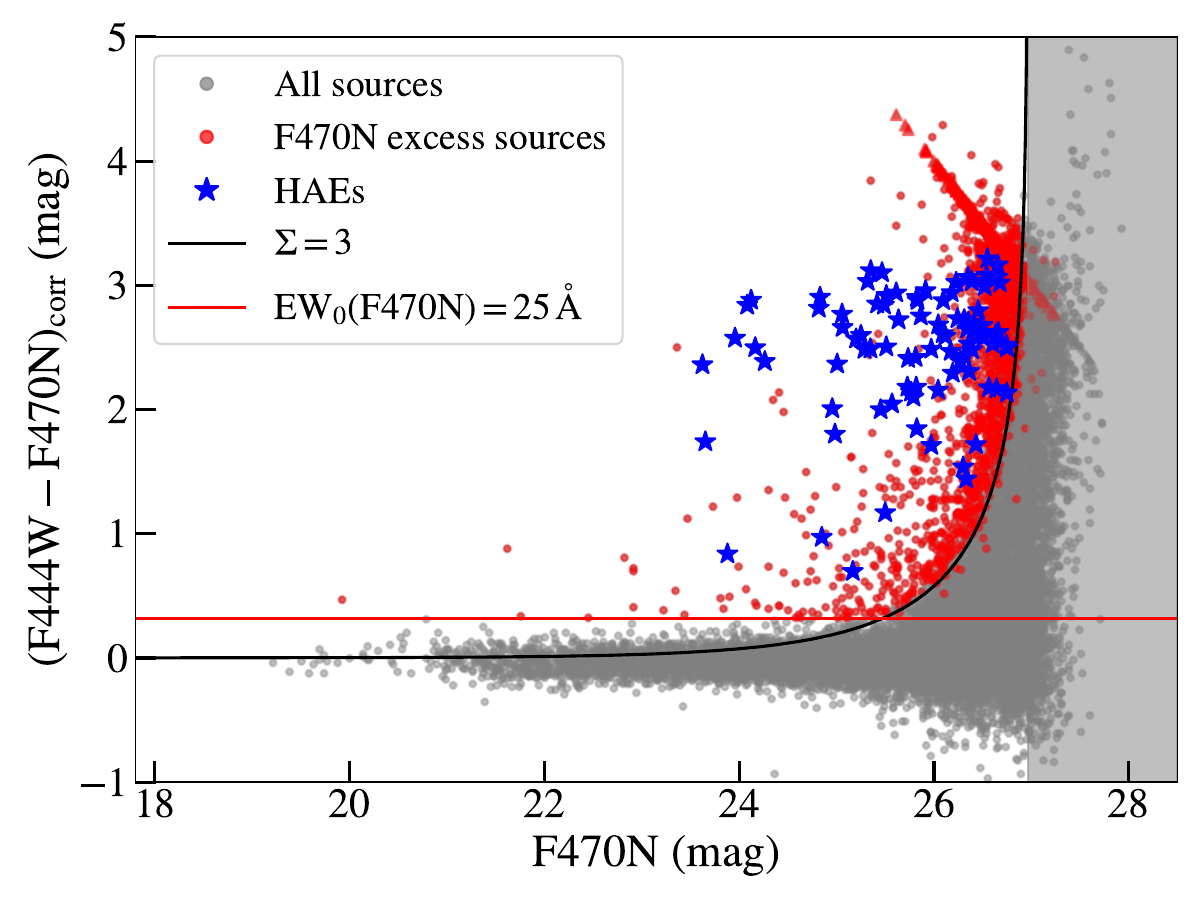}
    \caption{Colour--magnitude diagram of ${\rm F470N}$ versus $({\rm F444W}-{\rm F470N})_{\rm corr}$.
Grey dots represent all sources, while red points indicate the selected F470N excess sources.
For sources undetected in ${\rm F444W}$, lower limits are shown as upward-pointing triangles.
The red line corresponds to $\mathrm{EW}_0(\mathrm{F470N})>25$\,\AA ($({\rm F444W}-{\rm F470N})_{\rm corr}=0.32$).
The black curve shows the average $\Sigma=3$ threshold, and the grey shaded region marks the region fainter than the average $3\sigma$ limiting magnitude in ${\rm F470N}$.
Blue stars mark the selected HAEs. We note that some F470N excess sources appear below the average $\Sigma=3$ curve or within the grey shaded region because both are based on the average depth, whereas the selection is performed using local depth estimates for each individual source.}
    \label{fig:F470N_excess}
\end{figure}

After correcting for continuum-slope differences, we quantify the F470N excess significance following \citet{Pirie+25} as:
\begin{equation}
\Sigma=
\frac{1-10^{-0.4C}}
{10^{-0.4(\mathrm{ZP}-m_{\rm F470N})}\,\sqrt{\sigma_{\rm F470N}^2+\sigma_{\rm F444W}^2}},
\end{equation}
where $C \equiv ({\rm F444W}-{\rm F470N})_{\rm corr}$, $m_{\rm F470N}$ is the F470N magnitude, ZP is the zero point of F470N, and $\sigma_{\rm F470N}$ and $\sigma_{\rm F444W}$ are the $1\sigma$ flux uncertainties.
Thus, $\Sigma$ represents the continuum-corrected F470N excess flux divided by the combined F470N and F444W photometric uncertainties.
We compute the rest-frame equivalent width ($\mathrm{EW}_0$) following \citet{Pirie+25}, using the effective widths
$\Delta\lambda_{\rm F470N}$ and $\Delta\lambda_{\rm F444W}$ as:
\begin{equation} \label{eq:ew_selection}
\mathrm{EW}_0(\mathrm{F470N})
=
\frac{\Delta\lambda_{\rm F470N}\left(10^{0.4C}-1\right)}
{1-10^{0.4C}\,\Delta\lambda_{\rm F470N}/\Delta\lambda_{\rm F444W}}
\cdot\frac{1}{1+z},
\end{equation}
assuming that the emission line is narrow compared to the filter widths.
This is appropriate for star-forming galaxies, since the F470N effective width corresponds to $\simeq 3.3 \times 10^3~{\rm km~s^{-1}}$ at $z \simeq 6.2$, much broader than typical nebular lines (e.g. \citealt{deGraaff+24}). 
In Equation~\eqref{eq:ew_selection}, we adopt a redshift of $z=6.174$, for which the \Ha\ line falls at the centre of the F470N transmission (Figure~\ref{fig:filter_profiles}).
We select NB excess sources by requiring S/N$_{\mathrm{F470N}}>3$, $\Sigma>3$, and $C>0.32$, which corresponds to $\mathrm{EW}_0(\mathrm{F470N})>25$\,\AA.
Figure~\ref{fig:F470N_excess} presents the colour-magnitude diagram of F470N magnitude versus $({\rm F444W}-{\rm F470N})_{\rm corr}$. 
In total, we identify 1,141 (1,490) F470N excess sources with (without) F444W detections.

\subsubsection{Identification of HAEs at $z\simeq6.2$} \label{sec:select_HAE}
In addition to HAEs at $z\sim6.2$, the F470N excess sources also include emission-line galaxies such as Paschen-$\alpha$ emitters at $z\sim1.6$ or [O\,{\sc iii}] emitters at $z\sim8.4$.
To isolate HAEs, we estimate photometric redshifts (photo-$z$) using \textsc{eazy} \citep{EAZY}.
The \textsc{eazy} configuration file is based on that used for the JWST photometric catalogue in the DAWN JWST Archive (DJA; \citealt{Valentino+23})\footnote{\url{https://dawn-cph.github.io/dja/imaging/v7/}} and modified to suit our F470N excess catalogue.
The input photometry consists of $0^{\prime\prime}\!\!.6$-diameter aperture fluxes and their uncertainties in 12 \textit{HST} and \textit
{JWST} bands: \textit{HST}/ACS F435W, F606W, and F814W, and \textit{JWST}/NIRCam F090W, F115W, F150W, F200W, F277W, F356W, F410M, F444W, and F470N, as listed in Table~\ref{tab:HAE_filter} except NB872 filter.
We use the \texttt{sfhz/agn\_blue\_sfhz\_13} template set, which includes blue high-$z$ templates \citep{Larson+23}. 
We follow the DJA magnitude-prior setup, but evaluate the prior using F150W instead of \textit{HST}/F160W.
We evaluate the redshift probability distribution, $p(z)$, over $0.01 < z < 10$ with a grid spacing of $\Delta z = 0.01$, and define $z_{\rm med}$ as the median of this distribution and $p(z)_{\rm peak}$ as its peak value.
Following \citet{Pirie+25}, we select sources with photo-$z$ in the range of $5.5 < z_{\rm med} < 6.5$.
We further require robust redshift solutions by imposing $\chi^2/N_{\mathrm{filt}}\leq8$, $p(z)_{\mathrm{peak}} > 0.5$, and $\int^{6.5}_{5.5} p(z)\,dz>0.7$, where
$\chi^2$ and $N_{\mathrm{filt}}$ are the chi-squared value and the number of filters used in the fit.
Among the 1,141 F470N excess sources with F444W detection, 118 sources satisfy these criteria.
In contrast, none of the 1,490 F470N excess sources undetected in F444W meet these criteria.
We then perform a visual inspection of multi-band cutout images to remove spurious F470N detections caused by diffraction spikes from bright stars. 
We also reject sources with no clear counterpart at the F470N detection position when the source contributing the flux within the photometric aperture in the other bands has a centroid offset by more than $0^{\prime\prime}\!\!.2$ from the F470N centroid. 
These sources are excluded because the multi-band photometry used for the photo-$z$ estimate is likely associated with a different object rather than with the F470N detection.
Finally, we identify a total of 88 sources as HAEs at $z\simeq6.2$ (blue stars in Figure~\ref{fig:F470N_excess}).
To verify the spectroscopic redshifts of these 88 HAEs, we cross-match them with the latest spectroscopic data release (v4.4) from DJA \footnote{\url{https://dawn-cph.github.io/dja/spectroscopy/nirspec/}} \citep{Valentino+25,Pollock+25}.
As a result, 23 sources are matched, all of which are found to lie within the redshift range $z=6.1274\text{--}6.2021$, demonstrating that our selection criteria work well.
Additionally, our sample includes one HAE at $z=6.177$ confirmed by NIRCam/F356W slitless spectroscopy \citep{Backhaus+24}, bringing the total number of spectroscopically confirmed $z\simeq6.2$ HAEs to 24.

\subsubsection{Removal of AGN/LRD}

Our sample of HAEs may be contaminated by active galactic
nuclei (AGNs) and so-called Little Red Dots (LRDs; e.g. \citealt{Matthee+24}), which are very compact and red objects. 
While the physical nature of LRDs remains a subject of ongoing debate, recent studies suggest they likely host significant AGN activity (e.g. \citealt{Inayoshi+25}). 
Since our study focuses on the properties of star-forming galaxies and the escape of \Lya photons, it is crucial to remove these potential AGNs and LRDs from our sample.
First, we cross-match our catalogue with the Chandra X-ray data from the AEGIS-X Deep Survey \citep{Laird+09, Nandra+15} and, for sources for which spectroscopic data are available, check for broad \Ha\ or H$\beta$ components characteristic of broad-line AGNs, such as components with FWHM $\gtrsim 1000~{\rm km~s^{-1}}$ (e.g. \citealt{Harikane+23}).
As a result, we find no X-ray counterparts for any of our HAEs, and none of the spectroscopically confirmed sources show evidence for such broad components.
Next, we identify potential LRDs by applying the colour and compactness criteria defined by \citet{Labbe+25}. 
We find three HAEs satisfying these criteria and remove them from the sample.
Notably, one of these sources had already been reported as an LRD candidate by \citet{Kocevski+25}. 
The complete selection flow for the HAE sample is summarized in Table~\ref{tab:selection}.

\begin{table}
\centering
\caption{Selection criteria of HAEs at $z\simeq6.2$ and the number of sources after each step.}
\begin{tabular}{lc}
\hline    
Selection criteria & Number of sources \\ \hline
Initial catalogue & 19,836 \\
F470N excess selection & 2,631 \\
Photo-$z$ selection & 118 \\
Visual inspection & 88 \\
Removal of AGN/LRD & 85 \\ 
Removal of uncertain HAE candidate & 84 \\ \hline
\end{tabular}
\label{tab:selection}
\end{table}

\section{Method} \label{sec:method}
\subsection{SED fitting} \label{sec:HAE_SED_fitting}
For the SED fitting of the identified HAEs, we use a $0^{\prime\prime}\!\!.6$-diameter aperture, which is larger than the aperture used for the source detection, to ensure the recovery of the total flux, as also adopted in \citet{Pirie+25}. 
However, the larger aperture increases the risk of contamination from neighbouring sources.   
To mitigate this issue, we perform the following masked re-measurement.
The masks are generated using the PSF-matched F356W image, which is the deepest broad-band.
For each HAE, we extract a $10\arcsec \times 10\arcsec$ cutout, detect sources with \textsc{SExtractor}, and mask all surrounding sources except the target before remeasuring the fluxes.
We confirm that all HAEs satisfy the HAE selection criteria even after re-photometry with masking. 

We then derive physical properties by SED fitting to the multi-band photometry with \textsc{cigale} \citep{Boquein+19}.
We use the latest release, \textsc{cigale} \texttt{v2025.1}.
The input photometry comprises 12 bands (\textit{HST} + \textit{JWST}), except for NB872. 
To construct the model SEDs, we employ a delayed exponential star formation history (SFH), characterized by an e-folding time ($\tau_{\rm main}$) and the age of the onset of star formation. 
The stellar emission is modelled using the \citet{BruChar+03} simple stellar population (SSP) library, assuming a \citet{Chabrier+03} initial mass function (IMF).
We allow the stellar metallicity to vary across four sub-solar to solar values.
To model the dust attenuation, we apply a modified \citet{Calzetti+00} starburst attenuation law. 
The colour excess, $E(B-V)$, is treated as a free parameter, varying from 0.0 to 1.0, and is applied equally to the stellar continuum and nebular emission. 
This assumption is commonly adopted for high-$z$ galaxies without direct Balmer-decrement constraints and is supported by recent \textit{JWST} results suggesting smaller differential reddening at $z \gtrsim 5$ (e.g. \citealt{Karthikeyan+26}), although it remains subject to uncertainties related to dust geometry and galaxy properties.
For the redshift input, we adopt spectroscopic redshifts where available; otherwise, we fix the redshift to $z=6.174$.  
The complete set of input parameters and their explored grids are summarized in Table~\ref{tab:cigale_parameters}.
We note that the stellar and gas metallicities are only weakly constrained by the available photometric SEDs because of degeneracies with other parameters. 
We therefore perform a robustness check in which both metallicities are fixed to $0.2\,Z_\odot$, following \citet{Korber+26}. 
This value is close to the metallicity for most HAEs in the fiducial fits.
The resulting physical properties, including $E(B-V)$, are nearly unchanged, and none of our conclusions are affected. 
We therefore
adopt the fiducial fits with free metallicities in the following analysis.

\begin{table} 
\caption{Input parameters and explored grids for the \textsc{cigale} SED fitting.}
\begin{tabular}{llll}
\hline
Module           & Parameter      & range             & Scale       \\ \hline
SFH              & age            & 2--900 [Myr]   & Logarithmic \\
                 & $\tau_{\rm main}$            & 1--10000 [Myr] & Logarithmic \\ \hline
SSP              & metallicity     & 0.02--1 {[}$\mathrm{Z}_{\odot}${]}    & Logarithmic \\ \hline
Nebular emission & $\log U$           & $-4.0$\text{--}$-2.0$         & Linear      \\
                 & gas metallicity & 0.02--1 {[}$\mathrm{Z}_{\odot}${]}    & Logarithmic \\ \hline
Dust attenuation & $E(B-V)$         & 0.0--1.0           & Linear      \\
                 & power-law slope & $-2.0$\text{--}$0.5$          & Linear      \\ \hline
\end{tabular}
\label{tab:cigale_parameters}
\end{table}
One HAE candidate shows a distinctly poor SED fit ($\chi^2 \simeq 7$), while all others have $\chi^2 < 3$.
At $z=6.174$, the F814W filter covers rest-frame wavelengths of $982$--$1241\,\mathrm{\AA}$ at more than half of its peak transmission, largely blueward of the Ly$\alpha$ break.
The observed F814W flux of this candidate is substantially higher than predicted by the SED model, even when strong Ly$\alpha$ emission is considered.
The broad-band fluxes around [O\,{\sc iii}] and H$\alpha$ are also poorly reproduced by the model.
These discrepancies cast doubt on its identification as an HAE at $z\simeq6.2$, and we therefore exclude this source.
The final HAE sample consists of 84 galaxies, which is used as the parent HAE sample in the following analyses, as shown in Table~\ref{tab:selection}.  
For analyses of SED-derived physical properties (e.g. $M_*$ and $E(B-V)$), we further require ${\rm S/N}>2$ detections in at least six bands in total and at least two of the rest-frame UV bands (F090W, F115W, F150W, and F200W).
These criteria minimize the reliance on photometric upper limits and ensure basic constraints on the UV continuum shape.
Of the parent HAE sample, 21 HAEs do not satisfy these criteria and are excluded from the SED-based analyses only, leaving 
63 HAEs.
\begin{table*}
\centering
\caption{
Summary of the HAE subsamples used in this study.
The numbers in parentheses denote the number of HAEs
individually detected in NB872 at $>3\sigma$.
}
\label{tab:subsamples}
\renewcommand{\arraystretch}{1.15}
\begin{tabular}{@{}p{0.20\textwidth}cp{0.24\textwidth}p{0.38\textwidth}@{}}
\hline
Subsample & $N_{\rm HAE}$ & Main use & Selection criteria \\
\hline

Parent HAE sample
& 84
& Parent sample
& Final HAE sample after the selection steps in Table~\ref{tab:selection} \\

SED fitting sample
& 63
& SED-derived physical properties
& Parent HAEs with ${\rm S/N}>2$ in at least six bands, including at least two rest-frame UV bands \\

NB872 sample
& 56 (14)
& Stacking analysis and \Lya photometry
& Parent HAEs with reliable NB872 photometry \\

Individual-\fesca\ sample
& 40 (12)
& Relations between individual \fesca\ and galaxy properties
& HAEs included in both the SED fitting and NB872 samples \\

UV size sample
& 12 (7)
& Relations involving rest-frame UV size
& Individual-\fesca\ sample with F115W size constraints \\

Optical size sample
& 33 (12)
& Relations involving rest-frame optical size
& Individual-\fesca\ sample with F356W size constraints \\

\hline
\end{tabular}
\end{table*}

\begin{figure*}
    \centering
    \includegraphics[width=\textwidth]{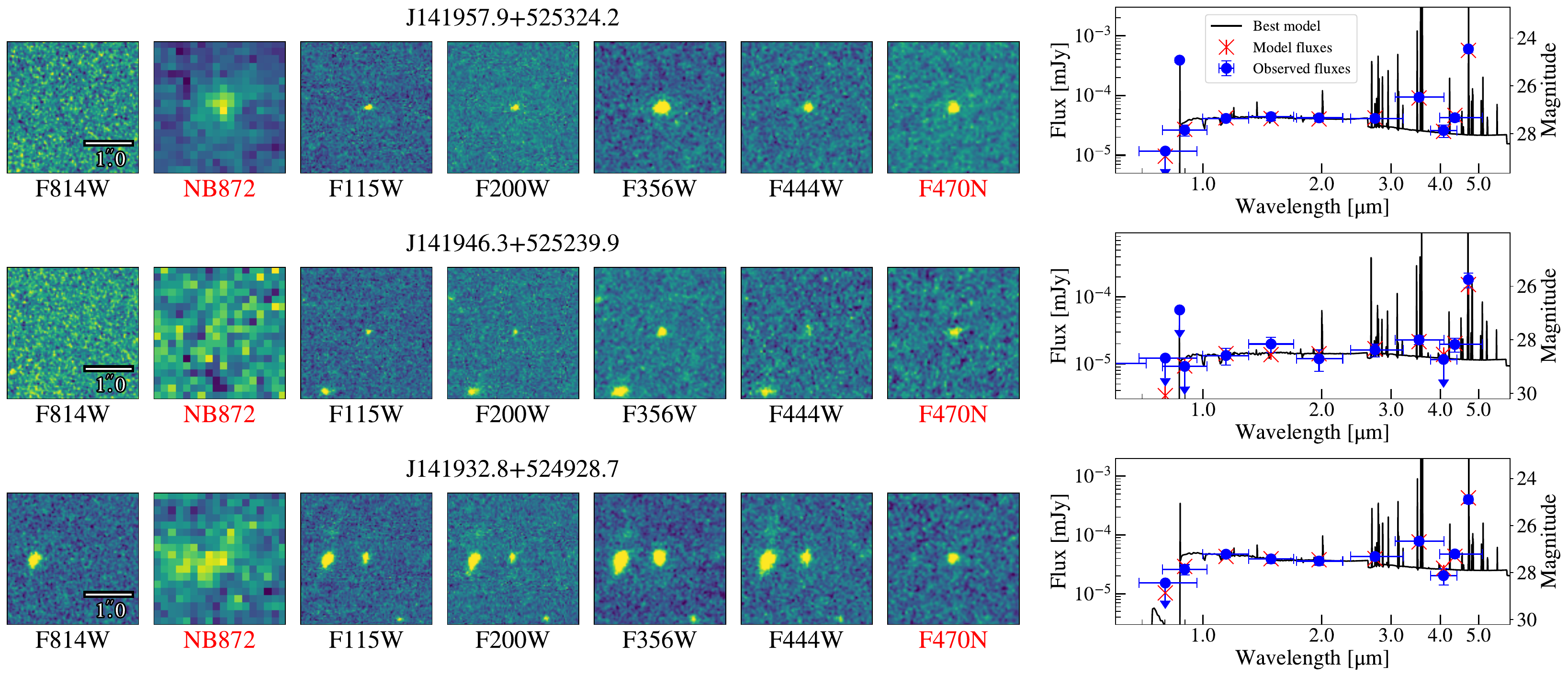}
    \caption{Examples of cutout images and SED fitting results for HAEs. 
    The rows show, from top to bottom, a Ly$\alpha$-detected HAE (NB872 $>3\sigma$), a Ly$\alpha$-undetected HAE (NB872 $\leq3\sigma$), and an HAE excluded from the NB872-based analyses because
    reliable NB872 photometry cannot be obtained due to contamination from a nearby source.
    The left-hand panels show $3\arcsec \times 3\arcsec $ cutout images, with the filter name indicated below each panel. The right-hand panels show the SEDs of the corresponding HAEs: the blue points represent the observed fluxes, the black line indicates the best-fitting SED model, and the red crosses mark the model-predicted fluxes in each filter.}
    \label{fig:HAE_cutout_SED}
\end{figure*}

\subsection{NB872 photometry }\label{sec:NB872_photometry}

We perform forced photometry for each of the 84 identified HAEs in a fixed $1^{\prime\prime}\!\!.5$-diameter aperture on the NB872 image centred at the F470N detection coordinates.
This aperture diameter corresponds to $\sim2.5$ times the NB872 PSF FWHM, or a physical radius of $\sim4.3$ pkpc at $z\simeq6.2$.
Previous studies have reported extended \Lya haloes with a median exponential scale length of $\sim4.5$ pkpc at $3<z<6$ \citep{Leclercq+17}, and scale lengths of $\sim5.9$ and $\sim12.6$ pkpc at $z=5.7$ and $6.6$, respectively \citep{Momose+14}.
Although this aperture may miss some extended diffuse \Lya emission, 
we adopt this to optimize the S/N while minimizing contamination from neighbours.

The photometric uncertainties are estimated from the error map produced by \texttt{hscPipe} and rescaled following the same empirical-noise procedure as in Section~\ref{sec:multi_catalog}. 
By comparing the error-map uncertainties with the scatter of randomly placed $1^{\prime\prime}\!\!.5$-diameter apertures in source-masked background regions, we find that the error map underestimates the aperture noise by a factor of 1.26. 
We therefore rescale the NB872 flux uncertainties by this factor, while retaining the spatial variation of the local noise.
Given the significantly coarser resolution of NB872 compared to \textit{HST}/\textit{JWST} images, contamination from nearby sources can significantly affect the aperture photometry, and we therefore apply the following masking procedure.
First, we construct a reference image by convolving the high-resolution \textit{JWST} F115W image to match the NB872 PSF and resampling it to the HSC pixel scale ($0^{\prime\prime}\!\!.168$) using the \texttt{reproject} package \citep{reproject}. 
Using this PSF-matched reference, we identify pixels within the photometric aperture whose flux is dominated by neighbouring sources and mask them on the NB872 image. 
We classify an HAE as unmeasurable when the PSF-matched F115W emission from a neighbouring source reaches the central region of the HAE ($r<2$ NB872 pixels from the F470N centroid, corresponding to about half of the NB872 PSF FWHM). 
In these cases, the expected central Ly$\alpha$-emitting region is contaminated, and reliable NB872
photometry cannot be obtained even after masking. 
These HAEs are therefore excluded from the NB872-based analyses.
Among the 84 HAEs, 28 are excluded due to such severe contamination, while 20 are retained after masking the contaminated pixels and 36 require no masking.
We confirm that this exclusion does not introduce a systematic bias in the HAE sample, as HAEs with and without reliable NB872 photometry have similar distributions of galaxy properties such as $M_*$, SFR, and $E(B-V)$.

Consequently, we obtain reliable NB872 photometry for 56 out of the 84 HAEs.
Among these, 14 sources are detected in NB872 at $>3\sigma$.
For the remaining 42 sources, we adopt $3\sigma$ upper limits in the subsequent analysis. 
When we restrict to the subsample used for SED-based analyses (Section~\ref{sec:HAE_SED_fitting}), we have reliable NB872 photometry for 40 sources, of which 12 are detected in NB872 at $>3\sigma$.
Several HAE subsamples used in the subsequent analyses and their selection criteria are summarized in Table~\ref{tab:subsamples}.
Figure~\ref{fig:HAE_cutout_SED} presents three examples of the HAE sample: a secure NB872 detection (NB872 $>3\sigma$), a source with valid photometry but below the detection threshold (NB872 $\leq3\sigma$), and a source excluded due to severe contamination.

\subsection{Measurement of the individual \Lya escape fractions} \label{sec:f_esc_Lya_calculation}
The F470N flux includes not only the H$\alpha$ line flux, but also the adjacent [N\,{\sc ii}]$\lambda\lambda$6548,6584 doublet flux, as well as the underlying continuum emission flux. 
To isolate the H$\alpha$ flux, we need to subtract these contaminating components.
Following \citet{Gimenez+25}, the continuum contribution is estimated using the best-fit SED models from \textsc{cigale}. 
The continuum flux density, $f_{\mathrm{cont,F470N}}$, is derived by convolving the F470N transmission curve over the continuum component of the model spectrum, excluding the contributions of the emission lines.
The total line flux, $F_{\mathrm{H\alpha + [N\,\text{\textsc{ii}}]}}$, is then calculated as:
\begin{equation}
F_{\mathrm{H\alpha + [N\,\text{\textsc{ii}}]}}
= (f_{\mathrm{F470N}} - f_{\mathrm{cont, F470N}})\times \Delta\lambda_{\mathrm{F470N}},
\end{equation}
where $f_{\mathrm{F470N}}$ is the observed flux density of the F470N filter.
The error assessment for the line flux takes into account the photometric errors of both the F470N and F444W.
Recent \textit{JWST} studies report relatively weaker [N\,{\sc ii}] flux compared to \Ha\ for galaxies at $z\sim3\text{--}7$, as $\log_{10}(\mathrm{[N\,\mbox{\textsc{ii}}]/H\alpha}) \sim -1.3$ to $-1.5$ (e.g. \citealt{Sanders+23, Shapley+23}). 
This corresponds to a flux correction by only $\sim 0.02$ dex; therefore, the [N\,{\sc ii}] contribution is considered negligible, following \citet{Pirie+25}.
We correct the H$\alpha$ flux for dust extinction using the colour excess $E(B-V)$ derived from the SED fitting, assuming the Calzetti law \citep{Calzetti+00}, and propagating the uncertainties in both quantities to obtain the intrinsic luminosity $L_{\mathrm{H\alpha,int}}$. 
The $\mathrm{SFR_{H\alpha}}$ is calculated using the following equation, based on the method of \citet{Kennicutt+98} adapted for \citet{Chabrier+03} IMF:
\begin{equation} 
    \log \mathrm{SFR_{H\alpha}}\,\mathrm{[M_{\odot} yr^{-1}]}=\log L_{\mathrm{H\alpha,int}}\,[\mathrm{erg/s}]-41.35.
\end{equation}

For the calculation of observed \Lya flux ($F_{\mathrm{Ly\alpha}}$), we use the NB872 magnitude ($m_{\mathrm{NB872}}$) and the 1216\,\AA\ continuum flux from the SED models ($F_{1216}$), following the method of \citet{Shimizu+25}: 
\begin{equation} \label{eq:shibuya}
    48.6+m_{\mathrm{NB872}}=-2.5\log_{10}\frac{\int^{\infty}_{0}[F_{1216}+F_{\mathrm{Ly\alpha}}\delta(\nu-\nu_{\mathrm{Ly\alpha}})]T_{\mathrm{NB872}}\mathrm{d\nu}}{\int^{\infty}_{0}T_{\mathrm{NB872}}\mathrm{d\nu}},
\end{equation}
where $T_{\mathrm{NB872}}$ is the transmission curve of the NB872 filter, and $\nu_{\mathrm{Ly\alpha}}$ is the observed frequency of the \Lya line.
We assume the \Lya line is approximated as a $\delta$-function, 
and $\nu_{\mathrm{Ly\alpha}}$ is taken to be the observed-frame Ly$\alpha$ frequency 
corresponding to the systemic redshift defined by H$\alpha$ in F470N, rather than the frequency at the centre of the NB872 filter.
IGM absorption affecting the continuum at wavelengths shorter than the \Lya line is incorporated using optical depths computed based on the \citet{Inoue+14} model.
$\rm EW_0(Ly\alpha)$ is derived by dividing $F_{\mathrm{Ly\alpha}}$ by $F_{1216}$.
For sources with spectroscopic redshifts, we calculate the $\rm EW_0(Ly\alpha)$ using their individual redshifts; for those without, we assume $z=6.174$ to compute the $\rm EW_0(Ly\alpha)$.

We derive \fesca\ by comparing the observed \Lya flux with the intrinsic \Lya flux inferred from the dust-corrected \Ha\ flux.
Assuming Case B recombination in gas with a temperature of $T_{\rm e}\sim10^4\,{\rm K}$ and an electron density of $n_{\rm e}\sim350\,{\rm cm^{-3}}$, for which the intrinsic \Lya/\Ha\ ratio is 8.7 \citep{Hummer},
\fesca\ is derived as:
\begin{equation}\label{eq:f_esc}
f_{\mathrm{esc}}^{\mathrm{Ly\alpha}}=\frac{F_{\mathrm{Ly\alpha,obs}}}{8.7F_{\mathrm{H\alpha,int}}}.
\end{equation}
Within Case B recombination, variations in gas conditions over $T_{\rm e}=5000$--$20000\,{\rm K}$ and $n_{\rm e}=100$--$1000\,{\rm cm^{-3}}$ change the intrinsic \Lya/\Ha\ ratio by only $\sim10\%$ \citep{Chen+24}.
In contrast, density-bounded H\,{\sc ii} regions or a porous ISM, as suggested by observations and simulations of \Lya and LyC escape (e.g. \citealt{Gazagnes+20,Kimm+22}), may approach Case-A-like conditions. 
In this case, the intrinsic \Lya/\Ha\ ratio can be as high as $\simeq12$, which would lower the inferred \fesca.
We nevertheless adopt the standard Case B value of 8.7 for consistency with previous studies.

\begin{figure}
    \centering
    \includegraphics[width=\columnwidth]{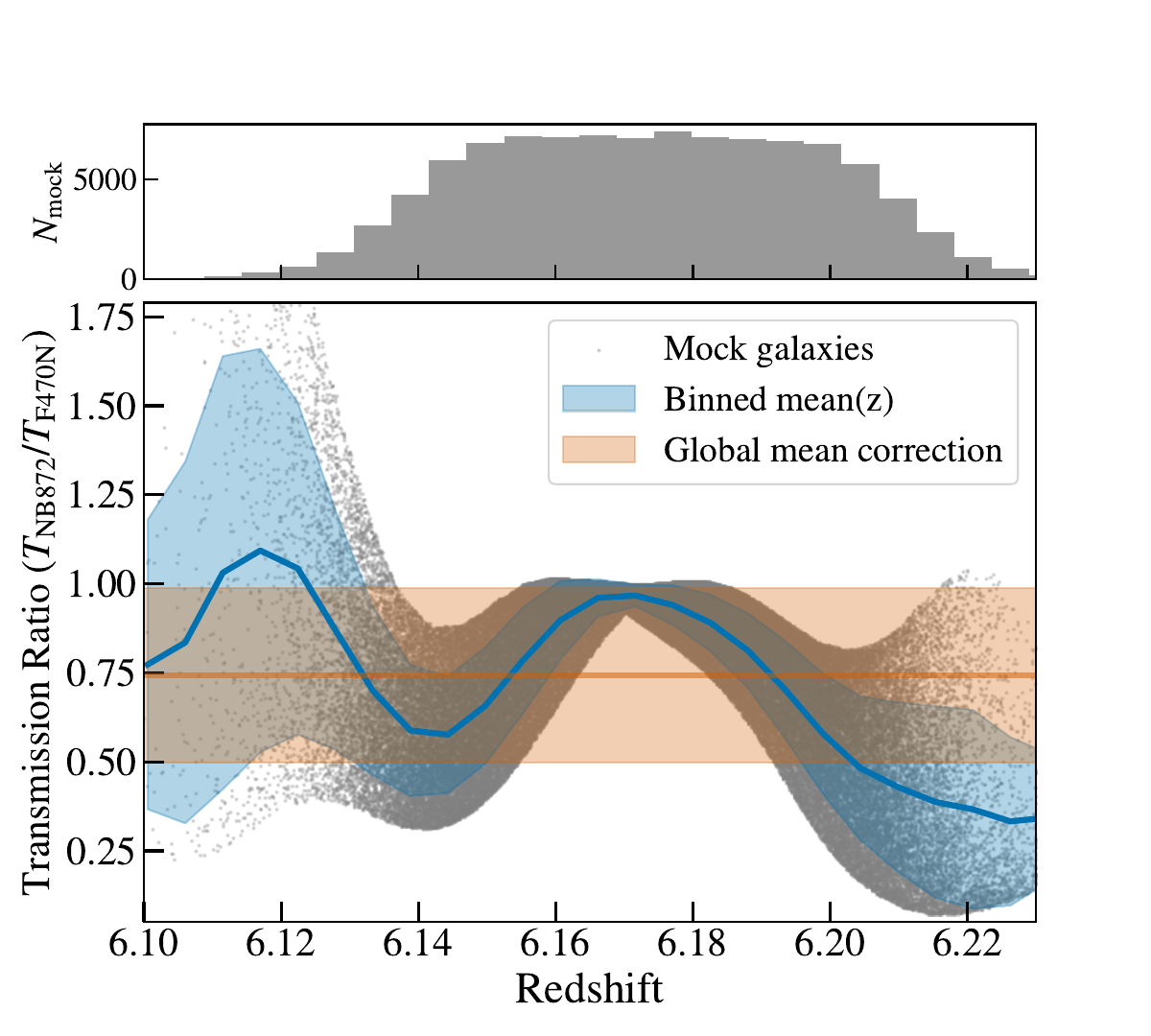}
    \caption{Transmission-ratio correction derived from Monte Carlo simulations. The upper panel shows the redshift distribution of the 100,000 mock HAEs. In the lower panel, grey points show 100,000 mock HAEs with redshifts drawn from the F470N transmission curve and Ly$\alpha$ velocity offsets randomly assigned in the range $0$--$800~\mathrm{km~s^{-1}}$. The y-axis is the filter transmission ratio $T_{\mathrm{NB872}}/T_{\mathrm{F470N}}$, where $T_{\mathrm{NB872}}$ is evaluated at the Ly$\alpha$ wavelength (after applying the velocity offset) and $T_{\mathrm{F470N}}$ at the H$\alpha$ wavelength. The blue curve denotes the binned mean as a function of redshift, with the shaded region indicating the $1\sigma$ scatter in each bin. The orange horizontal line and shaded region show the global mean transmission ratio and its $1\sigma$ scatter.}

    \label{fig:trans_corr}
\end{figure}

Comparing fluxes from distinct NB filters requires a correction for their differential transmission profiles, as discussed in \citet{Matthee+16}. 
As illustrated in Figure~\ref{fig:filter_profiles}, the NB872 transmission curve differs in shape and is slightly narrower than that of F470N. 
In addition, \Lya emission typically exhibits a positive velocity offset relative to the systemic redshift (defined here by \Ha) (e.g. \citealt{Tang+24a,Prieto-Lyon+25}). 
This spectral shift can move the \Lya line into a lower-transmission wing of the filter transmission, further biasing the \fesca measurement.
To quantify and correct for these effects, we perform a Monte Carlo simulation following the methodology of \citet{Nakajima+12} and \citet{Matthee+16}. 
We generate 100,000 mock HAEs that follow the redshift probability distribution defined by the F470N transmission curve, as shown in the top panel of Figure~\ref{fig:trans_corr}.
For each mock galaxy, we assign a Ly$\alpha$ velocity offset uniformly distributed over $0$--$800$ km s$^{-1}$ \citep{Endsley+22}, and compute the corresponding filter transmissions for Ly$\alpha$ (NB872) and H$\alpha$ (F470N).
In this way, we derive a transmission-based correction factor as a function of redshift as shown by the blue curve in Figure~\ref{fig:trans_corr}.
We correct $f_{\mathrm{esc}}^{\mathrm{Ly\alpha}}$ calculated from Equation~\eqref{eq:f_esc}
by dividing it by this ratio.
For the 17 HAEs with both spectroscopic redshifts and
reliable NB872 photometry, we adopt the redshift-dependent
mean ratio and its $1\sigma$ scatter. 
For the remaining 39 HAEs without spectroscopic redshifts, we adopt the global mean ratio, $0.723 \pm 0.256$, shown by the orange horizontal line.
The uncertainty in the adopted ratio is propagated into the final uncertainty of $f_{\mathrm{esc}}^{\mathrm{Ly\alpha}}$ for individual detections, with the relatively large scatter in the global ratio contributing substantially in addition to the NB872 photometric uncertainty.
We note that the \Lya flux used in Equation~\eqref{eq:f_esc} is the observed flux after attenuation by the ISM, CGM, and IGM. 
Therefore, the $f_{\rm esc}^{\rm Ly\alpha}$ derived in this work should be interpreted as an effective observed \Lya escape fraction, rather than an intrinsic escape fraction from the galaxy ISM alone.

\subsection{Size measurement} \label{sec:HAE_size_measure}
To characterize the spatial extent of the HAEs, we employ the two-dimensional surface brightness modelling tool \textsc{galight} \citep{GaLight}. 
For each band, we generate $2\arcsec \times 2\arcsec$ cutouts and model the galaxy morphology using a single S\'{e}rsic profile \citep{Sersic,Sersic_book} convolved with the empirical PSF for the corresponding band, constructed as described in Sections~\ref{sec:F470N_data} and \ref{sec:multi_catalog}. 
To prevent unphysical solutions, we constrain the S\'{e}rsic index, $n$, to the range $0.5 \le n \le 7$.
The fitting procedure consists of two steps: first, a global parameter search using Particle Swarm Optimization (PSO) to identify the maximum likelihood region, followed by Markov Chain Monte Carlo (MCMC) sampling to estimate the posterior distributions and uncertainties.
Previous studies indicate that structural parameters derived from 2D fitting codes, such as \textsc{galight} or \textsc{galfit} \citep{Galfit}, can become unreliable for low-S/N sources, particularly at S/N $\lesssim 10$
(e.g. \citealt{Allen+24}).
Consequently, we restrict our morphological analysis to bands with S/N $> 10$ in the $0^{\prime\prime}\!\!.6$-diameter aperture photometry.
The number of HAEs successfully modelled in each band is as follows: F090W: 6, F115W: 22, F150W: 15, F200W: 18, F277W: 33, F356W: 58, F444W: 31, and F470N: 20. 
Therefore, throughout the subsequent analysis, we use F115W (rest-frame $\sim$1600\,\AA) and F356W (rest-frame $\sim$5000\,\AA) for the rest-frame UV and optical sizes ($R_{\mathrm{e, UV}}$, $R_{\mathrm{e, opt}}$) respectively, because they provide the largest number of measured sources.
When we further restrict to the subsample used for SED-based analyses and with reliable NB872 photometry, the number of sources with size constraints reduces to 12 in the rest-frame UV and 33 in the restframe optical, as summarized in Table~\ref{tab:subsamples}.
To compare with other studies, we employ the semi-major axis radius output by \textsc{galight}
instead of the circularized radius.

Reliable size measurements can be challenging when the effective radius is small compared to the PSF, even for sources satisfying the S/N criterion, (e.g. \citealt{Stephenson+25}).
To assess the reliability of our measurements, we perform artificial-galaxy injection--recovery tests in the F115W and F356W images (e.g. \citealt{Ono+24,Lyu+25}).
We generate mock galaxies using single S\'{e}rsic profiles with $n=1$ and an axis ratio of $q=0.7$, convolve them with the corresponding empirical PSFs, and inject them into blank-sky regions of the CEERS F115W and F356W mosaics.
The source fluxes are scaled to yield an expected S/N of 10 in a $0^{\prime\prime}\!\!.6$-diameter aperture, 
corresponding to the S/N threshold adopted for our morphological analysis.
For each input radius, we generate 400 mock galaxies and recover their effective radii using the same \textsc{galight} fitting procedure as for the real HAEs.
We evaluate the size recovery using the median and 16th--84th percentile range of the recovered-to-input radius ratio, $R_{\mathrm{e,out}}/R_{\mathrm{e,in}}$.
In F115W, the recovery scatter decreases with increasing input radius and becomes approximately stable at $R_{\mathrm{e,in}}\gtrsim0^{\prime\prime}\!\!.021$.
At this radius, the 16th--84th percentile range is $0.907$--$1.058$.
We therefore conservatively adopt $R_{\mathrm{e,lim}}=0^{\prime\prime}\!\!.021$ (0.32 times the PSF FWHM) as the minimum reliably measurable radius in F115W.
Similarly, in F356W, the recovery scatter becomes approximately stable at $R_{\mathrm{e,in}}\gtrsim0^{\prime\prime}\!\!.075$.
We thus adopt $R_{\mathrm{e,lim}}=0^{\prime\prime}\!\!.075$ (0.54 times the PSF FWHM) as the empirical measurement floor for F356W.
For HAEs with fitted radii below these limits, we conservatively treat their sizes as upper limits.
The resulting surface densities are likewise treated as lower limits.
These limits are empirical thresholds calibrated for the fiducial mock-galaxy profiles, rather than source-specific statistical confidence bounds.

\subsection{Stacking analysis for the median Ly$\alpha$ escape fraction}
\subsubsection{Stacking procedure} \label{sec:stack_procedure}
\begin{figure}
    \centering
    \includegraphics[width=0.9\columnwidth]{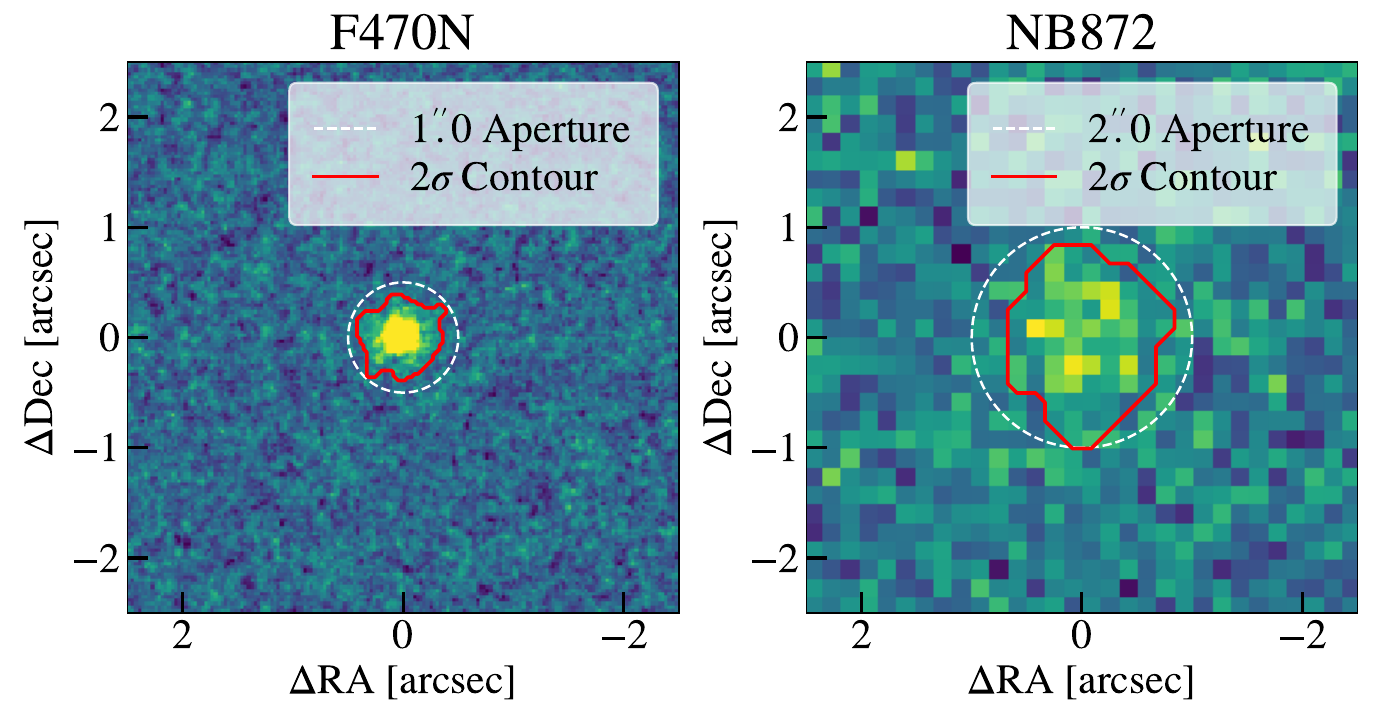}
    \caption{Stacked images of F470N (H$\alpha$ + optical continuum; left) and NB872 (Ly$\alpha$ + UV continuum; right) for HAEs at $z\simeq6.2$.
    Each image is centred on the stacked source with a size of $5\arcsec \times 5\arcsec$. The red contours indicate regions with S/N $> 2$, and the white dashed circle represents the aperture used for photometry.
    }
    \label{fig:stacked_images}
\end{figure}

We measure the \fescmed\ at $z\simeq6.2$ by generating deep H$\alpha$ and  Ly$\alpha$ images via stacking analysis of the NB data. 
Since the F470N and NB872 bands contain contributions from both emission lines and their underlying continua, we also perform stacking on the corresponding broad-band images (F115W for UV continuum; F444W for optical continuum). 
The stacking sample comprises the 56 HAEs for which reliable NB872 photometry can be obtained after masking contamination from nearby sources, irrespective of their individual NB872 detection significance.
First, we generate $20\arcsec \times 20\arcsec$ cutouts for each source in the NB872, F115W, F470N, and F444W bands.
We use the PSF-matched images of the F115W and F444W filters, homogenized to the F470N PSF (see Section \ref{sec:multi_catalog}). 
While Ly$\alpha$ emission is expected to be the most spatially extended component, observed Ly$\alpha$ haloes extend to at most $\sim 40$\,kpc ($\sim 7\arcsec$ at $z \sim 6.2$) \citep{Momose+14,Kusakabe+22}. 
Thus, our $20\arcsec$ window is sufficiently large to capture the full extent of the emission. 
To prevent contamination, we mask all sources other than the central target, following the procedure described in Sections \ref{sec:HAE_SED_fitting} and \ref{sec:NB872_photometry}.
We employ median stacking to combine the images.
This approach mitigates bias from extreme outliers or exceptionally bright sources and is consistent with previous studies \citep{Matthee+16}. 
Although we apply background subtraction to individual images prior to stacking (Section~\ref{sec:F470N_data} and Section~\ref{sec:NB872_data}), the stacked images still contain a residual background signal arising from faint, unresolved galaxies.
To correct for this, we define the region outside $7\arcsec$ from the centre as the background and subtract its median value uniformly from the image. 
We confirm that this background level is consistent with that derived from stacking random regions (including the sky regions).
Figure~\ref{fig:stacked_images} presents the resulting stacked images for F470N (H$\alpha$) and NB872 (Ly$\alpha$). 
For clarity, we display $5\arcsec \times 5\arcsec$ cutouts centred on the target. 
Red contours indicate regions where the surface brightness exceeds the $2\sigma$ detection threshold.

We then perform fixed circular aperture photometry on the stacked images.
For the F115W, F444W, and F470N stack images, which share a common PSF size, we employ a uniform aperture diameter of $1^{\prime\prime}\!\!.0$. 
As illustrated in Figure~\ref{fig:stacked_images}, this aperture captures the full extent of the $>2\sigma$ emission and exceeds the aperture-to-PSF ratio typically used for H$\alpha$ photometry \citep{Matthee+16}.
For the NB872 stack, the choice of aperture requires a careful balance between recovering the extended halo flux (e.g. \citealt{Momose+14}) and minimizing background noise.
Based on the analysis described in Section~\ref{sec:stacking_result}, we adopt a $2^{\prime\prime}\!\!.0$-diameter aperture, which fully encompasses the $>2\sigma$ Ly$\alpha$ emission.
Photometric uncertainties are estimated via bootstrap resampling. 
We generate 3,000 mock stacks by sampling with replacement from the parent stacking sample and repeat the photometry procedure for each stack. 
The 16th--84th percentile range of the resulting distribution is adopted as the $1\sigma$ confidence interval.
Using the stacked photometry, we derive the Ly$\alpha$ and H$\alpha$ luminosities following the methodology in Section \ref{sec:f_esc_Lya_calculation}. 
For the H$\alpha$ luminosity, we apply the dust correction using the
median value of $E(B-V)= 0.075$. 
Substituting these luminosities into Equation \eqref{eq:f_esc}, we calculate the \fescmed\ at $z \simeq 6.2$.

\subsubsection{Completeness weighting} \label{sec:completeness_weighting}
\begin{figure}
    \centering
    \includegraphics[width=0.85\columnwidth]{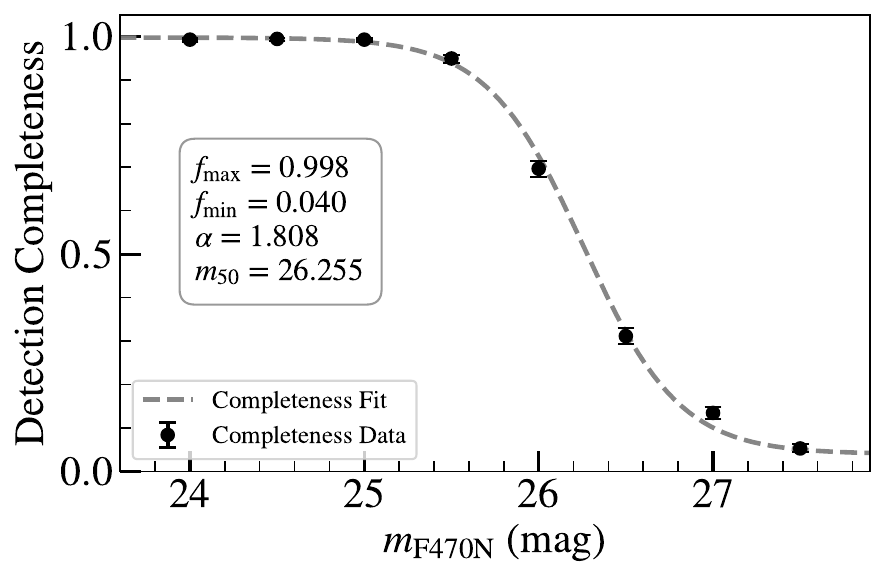}
    \caption{Detection completeness as a function of F470N magnitude, estimated from mock galaxy recovery simulations. The points represent the completeness in each magnitude bin, while the grey curve shows the best-fit function from \citet{Serjeant+00}.}
    \label{fig:detection_completeness}
\end{figure}

When comparing our stacked $f_{\mathrm{esc,med}}^{\mathrm{Ly\alpha}}$ with the \fesca\ based on LFs, it is necessary to clarify which luminosity range is effectively represented by the stack. 
This is because, as shown in \citet{Goovaerts+24b}, LF-based determinations of the \fesca\ could be sensitive to the faint-end integration limit.
Stacking only the sampled HAEs yields representative flux contributed by galaxies spanning the accessible H$\alpha$ luminosity range from the minimum \Ha\ luminosity set by the F470N limiting magnitude up to the bright end.
The faintest source in our sample has $L_{\rm H\alpha}=10^{41.3}\ {\rm erg\ s^{-1}}$ corresponding to $\sim 0.06\times L^*$, based on the H$\alpha$ LF at $z\simeq 6.15$ \citep{Covelo-Paz+25}, where 
the detection completeness is low.
Therefore, we apply a completeness correction in the stacking by estimating the detection completeness, $f_{\rm det}$, of each source from its $m_{\rm F470N}$ and weighting it by $w=1/f_{\rm det}$ in the weighted median stack.
The $f_{\rm det}$ is estimated via the recovery fraction of mock galaxies. 
We sample magnitudes in the range $m_{\rm F470N}=23.0$ to $28.0$ in steps of 0.5 mag, and embed 600 mock galaxies per magnitude bin into blank regions in the F470N image using the \textsc{Galsim} package \citep{GalSim}. 
The structural parameters of the mocks are chosen to span the same range of our HAEs measured in Section~\ref{sec:HAE_size_measure}.
We randomly assign S\'{e}rsic index $n=1$ to 2 and effective radii $R_{\rm e, F470N}=0.3$ to $0.9$ kpc, convolve the models with the PSF, and add them to the images. 
We then run \textsc{SExtractor} with the same configuration as in Section~\ref{sec:multi_catalog} and measure the $f_{\rm det}$ from the recovery fraction.
The resulting $f_{\rm det}$ as a function of $m_{\rm F470N}$ is shown in Figure~\ref{fig:detection_completeness}, where we observe a sharp decline around $m_{\rm F470N} \sim 26$.
We fit the $f_{\rm det}$ using the following function from \citet{Serjeant+00}:
\begin{equation}
    f_{\mathrm{det}}(m_{\rm F470N})=\frac{f_{\max}-f_{\min}}{2}\left\{\tanh\left[\alpha\left(m_{\rm F470N}^{50}-m_{\rm F470N}\right)\right]+1\right\}+f_{\min}.
\end{equation}
Here, $f_{\max}$ and $f_{\min}$ denote the upper and lower asymptotic completeness levels, respectively, $\alpha$ controls the steepness of the transition, and $m_{\rm F470N}^{50}$ represents the magnitude at which the $f_{\rm det}$ reaches the midpoint between these two limits. 
The best-fit parameters are $f_{\max}=0.998$, $f_{\min}=0.040$, $\alpha=1.808$, and $m_{\rm F470N}^{50}=26.255$. 
This function represents the completeness for each HAE based on its F470N magnitude, and the inverse of this completeness is applied as a weight in the stacking process.

\section{Results and Discussion} \label{sec:results_and_discussion}
\subsection{Physical properties of HAEs at $z\simeq6.2$} \label{sec:HAE_physical_properties}
We analyse the physical properties of the SED fitting sample, comprising 63 HAEs with robust SED constraints.
\begin{figure}
    \centering
    \includegraphics[width=\columnwidth]{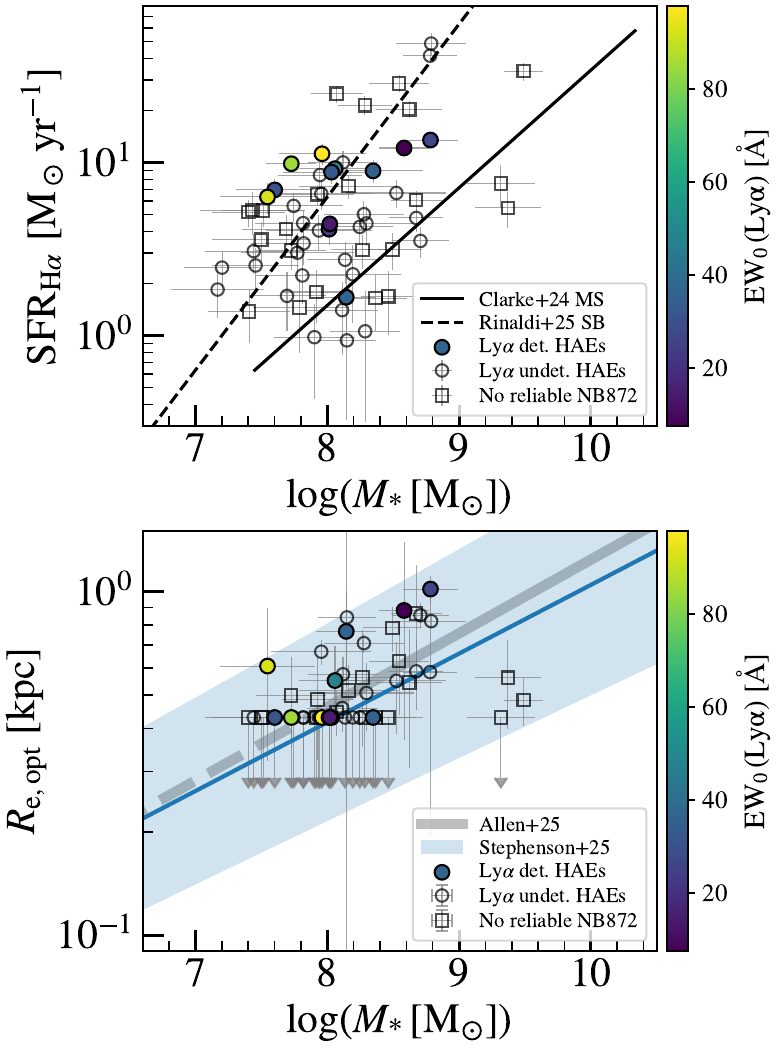}
    \caption{Top: The relationship between $\mathrm{SFR_{H\alpha}}$ and $M_*$ for HAEs at $z\simeq6.2$. The black solid line indicates the MS at $z\sim6$ from \citet{Clarke+24}, while the dashed line represents the SB region from \citet{Rinaldi+24}.
    Bottom: The relationship between $R_{e, \mathrm{opt}}$ and $M_*$. The grey line shows the size--$M_*$ relation from \citet{Allen+24}, and the blue shaded region corresponds to the relation from \citet{Stephenson+25}. 
    In both panels, filled circles denote Ly$\alpha$-detected HAEs (NB872 $>3\sigma$) and are colour-coded by their $\mathrm{EW}_0(\mathrm{Ly}\alpha)$, while open circles denote Ly$\alpha$-undetected HAEs and open squares denote HAEs without reliable NB872 photometry. Downward arrows in the bottom panel indicate upper limits on the optical effective radii for resolution-limited sources.}

    \label{fig:SFR_Ha_and_Re_vs_Mstar}
\end{figure}
The top panel of Figure~\ref{fig:SFR_Ha_and_Re_vs_Mstar} presents the relationship between SFR$_{\mathrm{H\alpha}}$ and $M_*$, while the bottom panel displays the size-$M_*$ relation, using the optical size, $R_{\mathrm{e, opt}}$, which serves as a robust proxy for the overall stellar distribution.
For sources with \Lya detections, the points are colour-coded by their \ewLya. 
Note that these plots include sources for which NB872 photometry is unavailable due to the contamination from nearby sources.
Regarding the SFR$_{\mathrm{H\alpha}}$--$M_*$ relation, the HAEs span a wide range, straddling both the Main Sequence (MS, \citealt{Clarke+24}) and the Starburst (SB) region \citep{Rinaldi+24}.
Notably, HAEs with larger \ewLya\ typically lie above the MS, indicating elevated star formation activity. 
Among the 12 Ly$\alpha$-detected HAEs shown in Figure~\ref{fig:SFR_Ha_and_Re_vs_Mstar}, 10 have $\rm EW_0(Ly\alpha)> 20$\,\AA, satisfying a commonly adopted EW criterion for \Lya emitters (LAEs; e.g. \citealt{Ouchi+20}).
The tendency for these HAEs to show elevated SFRs aligns with findings from previous studies (e.g. \citealt{Iani+24,Shimizu+25}), which demonstrate that LAEs generally exhibit higher SFRs compared to the typical MS population.
This trend remains unchanged even when the \Ha\ luminosities are not corrected for dust attenuation, indicating that it is not driven by the SED-based dust correction.

In the size--$M_*$ relation as shown in the bottom panel of Figure~\ref{fig:SFR_Ha_and_Re_vs_Mstar},  
the measured optical sizes overlap with the size--$M_*$ relations for general star-forming galaxies (SFGs) reported in \textit{JWST} observations (e.g. \citealt{Allen+24,Stephenson+25}). 
However, a substantial fraction of the optical sizes are resolution-limited and are therefore shown as upper limits following the injection--recovery tests described in Section~\ref{sec:HAE_size_measure}.
The present data thus do not allow us to establish a robust positive size--$M_\ast$ correlation and quantify its intrinsic scatter.
We also confirm that the $R_{\rm e,UV}$--$M_\ast$ relation, although based on a smaller subsample, is broadly consistent with that of \citet{Allen+24}.
Furthermore, we find no clear correlation between \ewLya\ and galaxy size, as LAEs lie on the same size--$M_*$ sequence as the general HAE population. 
This is consistent with previous studies (e.g. \citealt{Paulino-Afonso+18,Shimizu+25}), which show that the size difference between LAEs and general SFGs becomes small at $z \gtrsim 5$. 
Together with the observed increase in the LAE fraction toward $z \sim 6$ \citep{Kusakabe+20,Goovaerts+23}, this suggests that compact LAE-like morphologies become increasingly common among high-$z$ SFGs. 

\subsection{Median \Lya escape fraction at $z \simeq 6.2$ from stacking analysis} \label{sec:stacking_HAE}
\subsubsection{Median \Lya escape fraction} \label{sec:stacking_result}
\begin{figure}
    \centering
    \includegraphics[width=0.9\columnwidth]{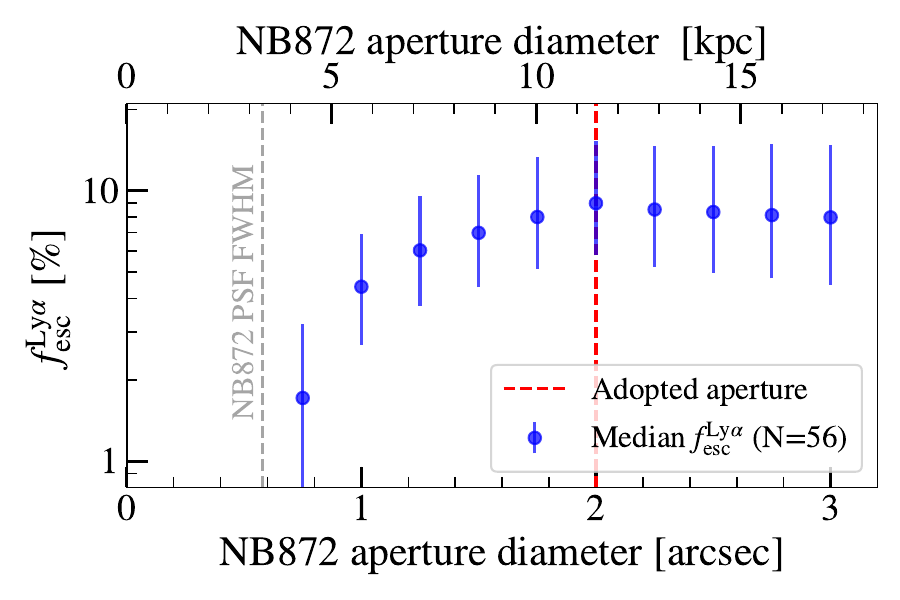}
    \caption{Dependence of the \fescmed\ on the aperture diameter used for photometry in the NB872 stacked image. The adopted aperture diameter of $2^{\prime\prime}\!\!.0$ is indicated by the vertical red dashed line.}
    \label{fig:nb872_aperture_photometry}
\end{figure}

Using the stacking procedure described in Section~\ref{sec:stack_procedure}, we first measure the $f_{\mathrm{esc,med}}^{\mathrm{Ly\alpha}}$ for the 56 HAEs with reliable NB872 photometry. 
From the unweighted stacks, we obtain $f_{\rm esc,med}^{\rm Ly\alpha} = 0.090^{+0.063}_{-0.032}$.
We test whether this result is sensitive to the aperture adopted for the NB872 stacked photometry. 
We repeat the NB872 aperture photometry by varying the aperture diameters ranging from $0^{\prime\prime}\!\!.75$ to $3^{\prime\prime}\!\!.0$ in steps of $0^{\prime\prime}\!\!.25$, while keeping the \Ha\ measurement fixed. 
Figure~\ref{fig:nb872_aperture_photometry} presents the resulting curve of growth. 
The derived $f_{\rm esc, med}^{\rm Ly\alpha}$ plateaus for aperture diameters larger than $2^{\prime\prime}\!\!.0$, justifying our adoption of the $2^{\prime\prime}\!\!.0$-diameter aperture in Section~\ref{sec:stack_procedure}.

We also compute a completeness-weighted stack to account for the decline in the F470N detection completeness at the faint end. 
Applying the inverse-completeness weights described in Section~\ref{sec:completeness_weighting}, we obtain $f_{\rm esc,med}^{\rm Ly\alpha} = 0.106^{+0.066}_{-0.044}$, which is consistent with the unweighted result within the uncertainties.
The weighted stack is nevertheless useful for comparison with LF-based estimates, as it corrects for the declining F470N detection completeness relative to the unweighted median stack. 
However, stacking- and LF-based approaches do not necessarily trace identical quantities, and such comparisons should therefore be interpreted with caution.

\subsubsection{Redshift evolution and comparison with previous studies}
\begin{figure*}
    \centering
    \includegraphics[width=0.7\textwidth]{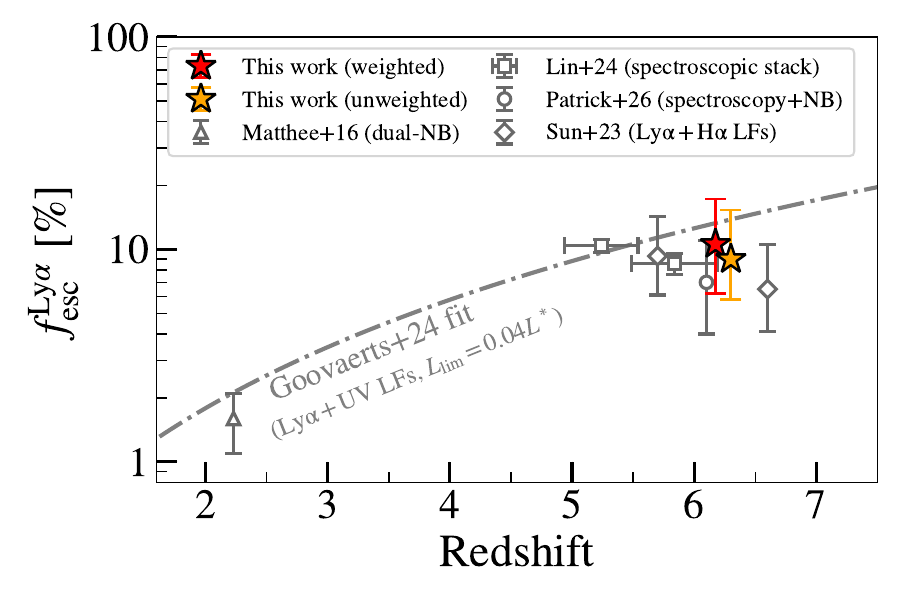}
    \caption{Redshift evolution of the \fescmed\ compared to previous studies. We plot our completeness-weighted (red stars) and unweighted (orange stars) results, with the unweighted points slightly shifted horizontally for visibility. Grey symbols show literature data from stacking methods (triangles: \citealt{Matthee+16}; squares: \citealt{Lin+24}), survival analysis (circles: \citealt{Patrick+26}), and LF comparisons (diamonds: \citealt{Sun+23}). The grey dash-dot curve shows the evolution of the \fesca\ derived from \Lya and UV LFs with an integration limit of $0.04L^*$ \citep{Goovaerts+24b}.}
    \label{fig:compare_f_esc_Lya_vs_z}
\end{figure*}

Figure~\ref{fig:compare_f_esc_Lya_vs_z} compares our results with previous measurements of \fesca\ across cosmic time. 
We include: $z \sim 2.2$ measurements using the analogous dual-NB technique \citep{Matthee+16}; estimates based on H$\alpha$ and Ly$\alpha$ LFs \citep{Sun+23}; spectroscopic stacking of HAEs at $z = 4.9\text{--}6.3$ \citep{Lin+24}; and LF-based measurements combining Ly$\alpha$ and UV data (\citealt{Goovaerts+24b}, grey dash-dot line). 
Our measurement at $z \simeq 6.2$ shows a significant increase of more than an order of magnitude compared to that at $z \sim 2.2$ obtained with the same dual-NB technique \citep{Matthee+16}, and is consistent with recent estimates at $z \sim 6$ \citep{Sun+23,Lin+24}.
More recently, \citet{Patrick+26} investigated an \Ha-selected sample at $z\simeq6.1$ using \textit{JWST}/NIRCam NB imaging and VLT/MUSE
\Lya spectroscopy.
They obtained a reverse Kaplan--Meier median $f_{\mathrm{esc}}^{\mathrm{Ly\alpha}}$ of $0.07^{+0.04}_{-0.03}$, consistent with our median-stack results within the uncertainties.
This result supports the redshift evolution in which the \fesca\ increases toward higher redshift, as previously suggested from comparisons between \Lya and UV LFs \citep[e.g.][]{Hayes+11,Konno+16,Goovaerts+24b}. 
Such an increase may reflect the fact that galaxies at $z\sim6$ are, on average, less dust-obscured and have bluer UV slopes than those at lower redshift \citep[e.g.][]{Bouwens+14,Topping+24}, and are also expected to be relatively metal-poor \citep[e.g.][]{Nakajima+23}. 
In addition, 
previous theoretical studies suggest that the gas and dust distributions in high-$z$ galaxies may be more clumpy or patchy, producing low-column-density sightlines of \Hi\ and dust through which Ly$\alpha$ photons can more efficiently escape \citep{Popping+17,Narayanan+18}, as also discussed in Section~\ref{sec:corr_fesca}.
At the population level, this interpretation is also consistent with \citet{Shimizu+25}, who argued that the apparent redshift evolution of the cosmic-averaged \fesca\ of SFGs is driven largely by an increasing fraction of LAE-like galaxies within the SFG population.

\subsubsection{Dependence on the luminosity limit}
\begin{figure}
    \centering
    \includegraphics[width=0.85\columnwidth]{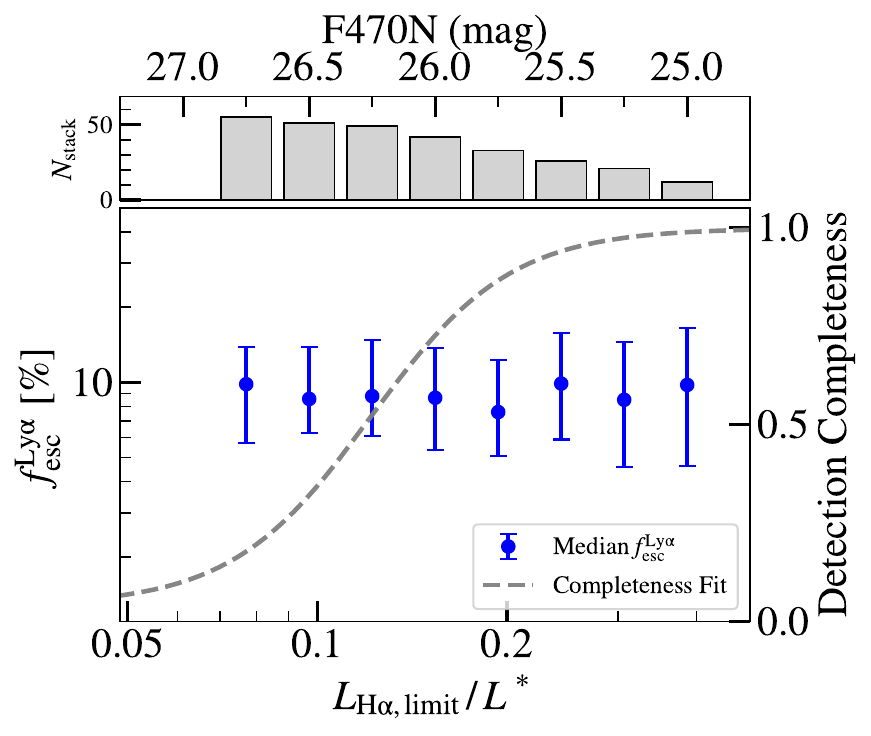}
    \caption{Dependence of the \fescmed\ on the lower luminosity limit used for stacking the HAE sample. The bottom axis shows the luminosity ratio relative to the characteristic luminosity $L^*$ of the H$\alpha$ luminosity function at $z\sim6.15$ \citep{Covelo-Paz+25}, while the top axis indicates the corresponding F470N magnitude. The upper histogram shows the number of HAEs included in each stack. The left axis gives \fescmed\ measured from stacks of HAEs brighter than each limit, while the right axis shows the F470N detection completeness estimated from mock-galaxy recovery simulations. The grey dashed curve shows the best-fit completeness function.}
    \label{fig:f_esc_L_dependense}
\end{figure}

To examine whether \fescmed\ is sensitive to the \Ha\ luminosity range included in the stack, we repeat the stacking analysis after imposing different lower limits on $L_{\rm H\alpha}$. 
Figure~\ref{fig:f_esc_L_dependense} shows the resulting \fescmed\ as a function of the adopted lower luminosity limit.
Even if the luminosity limit of the sample is changed, no significant change in \fesca\ is observed.
This trend is also observed 
even when the stack is restricted to luminosity limits where the F470N detection completeness is high, suggesting that the uncertainty in completeness correction does not play a significant role. 
This result contrasts with LF-based estimates, in which the inferred \fesca\ depend on the adopted faint-end integration limit (e.g. \citealt{Goovaerts+24b}).
In contrast to LF-based approaches, our stacking measurement uses \Ha-selected galaxies within the luminosity range directly probed by the observations. 
Although we correct for the F470N detection completeness, the measurement does not rely on extrapolating LFs to the unobserved faint end or on abundance-matching-like assumptions that relate different galaxy populations selected by UV, \Lya, or \Ha\ emission.

On the other hand, the absence of a variation with the adopted lower luminosity limit may simply reflect the limited statistical precision of the present stacked measurements, which is determined by both the photometric uncertainties of individual galaxies and the number of sources included in each stack.
Following the method of \citet{Goovaerts+24b}, we estimate the \fesca\ from the \Lya and UV LFs for the case of $L_{\rm limit} = 0.25\,L^\ast$. 
Compared to the case of $L_{\rm limit} = 0.04\,L^\ast$, this estimate at $z \sim 6$ should be lower by about $0.2$ dex for $L_{\rm limit} = 0.25\,L^\ast$. 
The uncertainty of our $f_{\mathrm{esc}}^{\mathrm{Ly\alpha}}$ measurements is comparable to this expected difference.
Therefore, to robustly investigate the \Ha-luminosity dependence of directly measured stacked \fescmed\, rather than relying on LF-based estimates, 
larger and deeper samples are needed through wider-field observations and deeper observations, including those of gravitationally lensed fields.

\subsubsection{Implications for LyC escape and the ionizing photon budget}
The weak variation of \fescmed\ with the adopted lower \Ha\ luminosity limit may have implications for Lyman-continuum (LyC) leakage. 
This is because both \Lya and LyC escape are sensitive to the distribution and covering fraction of \Hi, and low-column-density channels in the ISM and CGM are expected to facilitate the escape of ionizing photons as well as \Lya photons (e.g. \citealt{Gazagnes+20,Kimm+22}). 
In addition, both observations and simulations suggest a good correlation between $f_{\mathrm{esc}}^{\mathrm{Ly\alpha}}$ and LyC escape fraction (\fescc) (e.g. \citealt{Begley+24,Choustikov+24c}). 
However, this comparison should be interpreted with caution. 
\Lya photons can escape after being shifted out of resonance by gas kinematics, whereas LyC photons require genuinely optically thin \Hi\ sightlines, and as a result, \Lya escape may not always fully trace LyC escape. 
Another important caveat is that direct LyC measurements determine only the line-of-sight escape fraction, while the quantity relevant for the ionizing photon budget is the angle- and time-averaged escape of LyC photons. 
Radiation-hydrodynamic simulations show that LyC escape can be highly stochastic and anisotropic, regulated by stellar feedback, the opening and closing of low-column-density channels, and viewing-angle effects (e.g. \citealt{Wise+09,Trebitsch+17,Rosdahl+22}). 
Therefore, even if no clear luminosity dependence is observed in line-of-sight LyC escape measurements, the intrinsic $f_{\mathrm{esc}}^{\mathrm{LyC}}$ relevant for the comoving ionizing emissivity  (\nion) that reaches the IGM could depend on luminosity. 

Observationally, the luminosity dependence of \fescc\ is still under debate, and most constraints have been discussed in terms of UV luminosity.
For example, \citet{Papovich+26} inferred the \fescc\ from fitting nebular emission lines and stellar continua measured from \textit{JWST} spectroscopy, and found no clear dependence on UV luminosity.
While some direct LyC observations also report no clear luminosity dependence (e.g. \citealt{Marchi+17,Bian+20}), others have found a significant dependence with higher escape fractions in fainter galaxies (e.g. \citealt{Steidel+18,Begley+22}).
\citet{Simmonds+24a} showed that models in which \fescc\ increases toward fainter UV magnitudes enhance 
the relative contribution of faint UV galaxies to the \nion, compared to models with a fixed \fescc\ across UV magnitudes.
Although this comparison is not directly equivalent to our \Ha-based result, it illustrates that the assumed luminosity dependence of \fescc\ can strongly affect which galaxy population dominates the ionizing photon budget during the EoR.

We therefore consider the implication of our result more directly in terms of \Ha\ luminosity. 
Following \citet{Korber+26}, \nion\ can be expressed as
\begin{align}
    \dot{n}_{\rm ion}=\int f_{\rm esc}^{\rm LyC}Q_{\rm ion}\phi_{\rm H\alpha}\,\mathrm{d}L_{\rm H\alpha},
\end{align}
where $Q_{\rm ion}$ is the ionizing photon production rate, and $\phi_{\rm H\alpha}$ is the H$\alpha$ LF.
$Q_{\rm ion}$ can be expressed in terms of the H$\alpha$ luminosity through a conversion factor $C_{\alpha}$ as $Q_{\rm ion}=L_{\rm H\alpha}/C_{\alpha}(1-f_{\rm esc}^{\rm LyC})$.
If \fescc is independent of H$\alpha$ luminosity, and the H$\alpha$ LF can be described by a Schechter function \citep{Schechter} with normalization density $\phi^*$, characteristic luminosity $L^*$, and faint-end slope $\alpha$, the contribution to \nion\ per logarithmic luminosity interval can be expressed as
\begin{align} \label{eq:nion_contribution}
    \frac{\mathrm{d} \dot{n}_{\rm ion}}{\mathrm{d} (\log L_{\rm H\alpha})} &= \frac{f_{\rm esc}^{\rm LyC} L_{\rm H\alpha}}{C_{\alpha}(1-f_{\rm esc}^{\rm LyC})} \phi^* \left(\frac{L_{\rm H\alpha}}{L^*_{\rm H\alpha}}\right)^{\alpha+1} \exp\left(-\frac{L_{\rm H\alpha}}{L^*_{\rm H\alpha}}\right)\notag  \\
    &\propto L_{\rm H\alpha}^{\alpha+2} \exp\left(-\frac{L_{\rm H\alpha}}{L^*_{\rm H\alpha}}\right).
\end{align}
As recent JWST observations have shown (e.g. \citealt{Covelo-Paz+25,Fu+25}), the faint-end slope $\alpha$ of the H$\alpha$ LF evolves relatively weakly with redshift and is approximately $\alpha \sim -1.6$ at $z \sim 6.15$, which is shallower than the faint-end slope of the UV LF.
In Equation~\eqref{eq:nion_contribution}, if $\alpha > -2$, the contribution to \nion\ per logarithmic H$\alpha$ luminosity interval peaks at a characteristic H$\alpha$ luminosity of $L_{\rm H\alpha}\sim(\alpha+2)L^*_{\rm H\alpha}$.
For example, with $\alpha = -1.6$, the H$\alpha$ luminosity interval that contributes most to \nion\ is around $0.4L^*_{\rm H\alpha}$, which lies within the high-completeness range of our F470N observations (see Figure~\ref{fig:f_esc_L_dependense}).
Therefore, if \fescc\ is approximately independent of \Ha\ luminosity, the ionizing photon budget is expected to be dominated not by the faintest \Ha\ emitters, but rather by 
galaxies with relatively high \Ha\ luminosities
closer to the observable regime, as also discussed by \citet{Korber+26}.
We note, however, that this argument concerns the contribution as a function of \Ha\ luminosity. 
The corresponding contribution as a function of UV luminosity depends on the relation between \Ha\ and 
UV luminosities, and is therefore not addressed here.

\subsection{Correlation between galaxy properties and the Ly$\alpha$ escape fraction} \label{sec:corr_fesca}
\begin{figure*}
    \centering
    \includegraphics[width=0.8\textwidth]{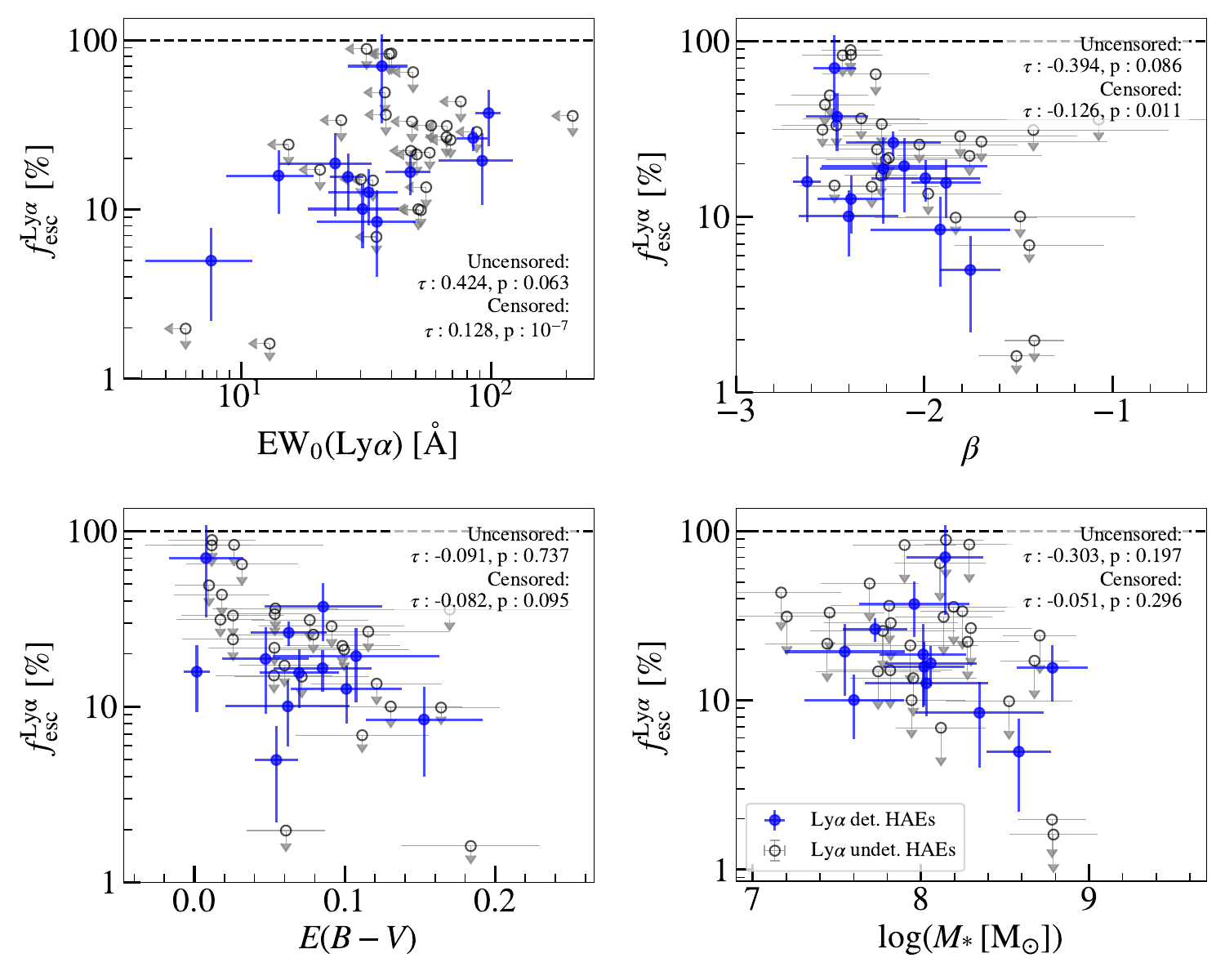}
    \caption{The relationships between \fesca\ and key physical properties for 40 HAEs at $z\simeq6.2$: \ewLya, UV slope $\beta$, $E(B-V)$, and $M_*$. In each panel, filled blue circles represent Ly$\alpha$-detected HAEs (NB872 $>3\sigma$), while open grey circles indicate Ly$\alpha$-undetected HAEs. Downward arrows indicate $3\sigma$ upper limits on \fesca, while leftward arrows indicate upper limits on \ewLya. Error bars indicate $1\sigma$ uncertainties. The Kendall rank correlation coefficients $\tau$ and $p$-values are shown for two cases: `Uncensored' (using measurements without limits in either variable) and `Censored' (including limits in either variable).}

    \label{fig:f_esc_vs_physical_properties}
\end{figure*}

We investigate the origin of the galaxy-to-galaxy variation in \fesca using individual measurements for the subsample of 40 HAEs with both reliable NB872 photometry and robust SED constraints, as summarized in Table~\ref{tab:subsamples}.
Figure~\ref{fig:f_esc_vs_physical_properties} presents the \fesca\ as a function of key physical properties: \ewLya, UV slope $\beta$, $E(B-V)$, 
and $M_*$.
To quantify correlations, we compute the Kendall rank correlation coefficient, $\tau$, and assess significance using two-sided $p$-values. 
The \fesca\ and \ewLya\ of 40 HAEs include upper limits for Ly$\alpha$-undetected HAEs.
We therefore treat Ly$\alpha$-related quantities as left-censored data and apply the generalized (censored) Kendall $\tau$ test for bivariate data with censoring (e.g. \citealt{Isobe+86,Akritas+96}).
Ly$\alpha$ and LyC escape fraction studies commonly involve non-detections and upper limits, making statistical methods that handle censored data essential (e.g. \citealt{Lin+24,Mascia+25}). 
Accordingly, we adopt the censored Kendall approach to incorporate upper limits in our correlation analysis.

Each panel of Figure~\ref{fig:f_esc_vs_physical_properties} shows Kendall's rank correlation coefficient $\tau$ and the $p$-value measured using detections only (uncensored), as well as $\tau$ and the $p$-value measured using censored data, respectively. 
First, for the relation between \ewLya\ and the \fesca, a statistically significant positive correlation appears, indicating that galaxies with stronger \Lya emission show more efficient \Lya escape. 
Multiple studies also report this trend (e.g. \citealt{Sobral+19,Begley+24,Napolitano+24}).

The UV slope $\beta$ also shows a statistically significant negative correlation with the \fesca\ for
both the detections-only case and 
the censored-data case. 
This correlation, where bluer $\beta$ corresponds to a larger \fesca\, has also been reported in the previous studies \citep{Matthee+16,Lin+24,Napolitano+24, Ning+26}, 
suggesting that lower dust attenuation makes \Lya escape easier. 
However, for $E(B-V)$, no statistically significant correlation with the \fesca\ appears.
The key point here is that $\beta$ and $E(B-V)$ do not necessarily represent the same quantities. 
The UV slope $\beta$ is weighted toward the least obscured UV-bright star-forming components that are visible to the observer, whereas the SED-derived $E(B-V)$ inferred from integrated photometry represents a luminosity-weighted average attenuation across the galaxy.
In vigorously star-forming galaxies at high-$z$, complex star-dust geometries can lead to scenarios where, despite a high overall dust content, optically thin sightlines remain, resulting in a blue $\beta$ as demonstrated by radiative transfer simulations \citep{Popping+17,Narayanan+18}.
\Lya is a resonant scattering line, making it more susceptible to dust attenuation than the continuum due to the increased path length from scattering (e.g. \citealt{Verhamme+06}). 
Therefore, our results suggest that \Lya escape is more strongly linked to the properties of the low-attenuation component reaching the observer (as traced by $\beta$) rather than the galaxy-averaged attenuation ($E(B-V)$).
The physical picture in which local low-attenuation regions regulate \Lya escape is also supported by recent spatially resolved \textit{JWST} observations of a gravitationally lensed LAE at $z=6.57$, where \Lya emission is found to emerge predominantly from a young, dust-cleared clump embedded in a multiphase ISM \citep{Markov+26}.
Such results suggest that local feedback and small-scale dust/gas geometry can regulate \Lya\ escape even when the galaxy as a whole is moderately dusty.
We note that the lack of a significant correlation with $E(B-V)$ could also be influenced by various factors,
such as uncertainties in the SED-derived $E(B-V)$, the limited sample size, and the metallicity
dependence of the dust content. 
Although the metallicities of individual HAEs in our sample are not tightly constrained by our SED fitting, high-$z$ low-mass galaxies are typically metal-poor, with recent JWST/NIRSpec studies suggesting $12+\log({\rm O/H})\sim7.7$--8.0 for galaxies with
$M_\ast\sim10^8$--$10^9\,M_\odot$ (e.g. \citealt{Nakajima+23,Curti+24}). 
At such low metallicities, the dust-to-gas ratio is expected to decrease rapidly, as indicated by
the steepening of the dust-to-gas--metallicity relation below $12+\log({\rm O/H})\sim8.0$ \citep{Remy-Ruyer+14,DeVis+19}. 
In this regime, the dynamic range of the
SED-derived $E(B-V)$ 
may narrow, potentially reducing the reliability of the constraints.
These effects, together with the limited sample
size, may dilute an intrinsic dependence of $f_{\mathrm{esc}}^{\mathrm{Ly\alpha}}$ on $E(B-V)$.

No significant correlation is found between
$M_*$ and \fesca.
We further examine the relationships between $f_{\mathrm{esc}}^{\mathrm{Ly\alpha}}$ and $\mathrm{SFR}_{\mathrm{H\alpha}}$, $\mathrm{sSFR}_{\mathrm{H\alpha}}$, $\mathrm{SFR}_{\mathrm{UV}}$, and $M_{\mathrm{UV}}$, as shown in Figure~\ref{fig:f_esc_vs_add} in Appendix~\ref{app:additional_correlations}.
No statistically significant correlations are found for any of these quantities.
It cannot be ruled out that the lack of significant correlations in these quantities is due to the small sample size.

\begin{figure}
    \centering
    \includegraphics[width=\columnwidth]{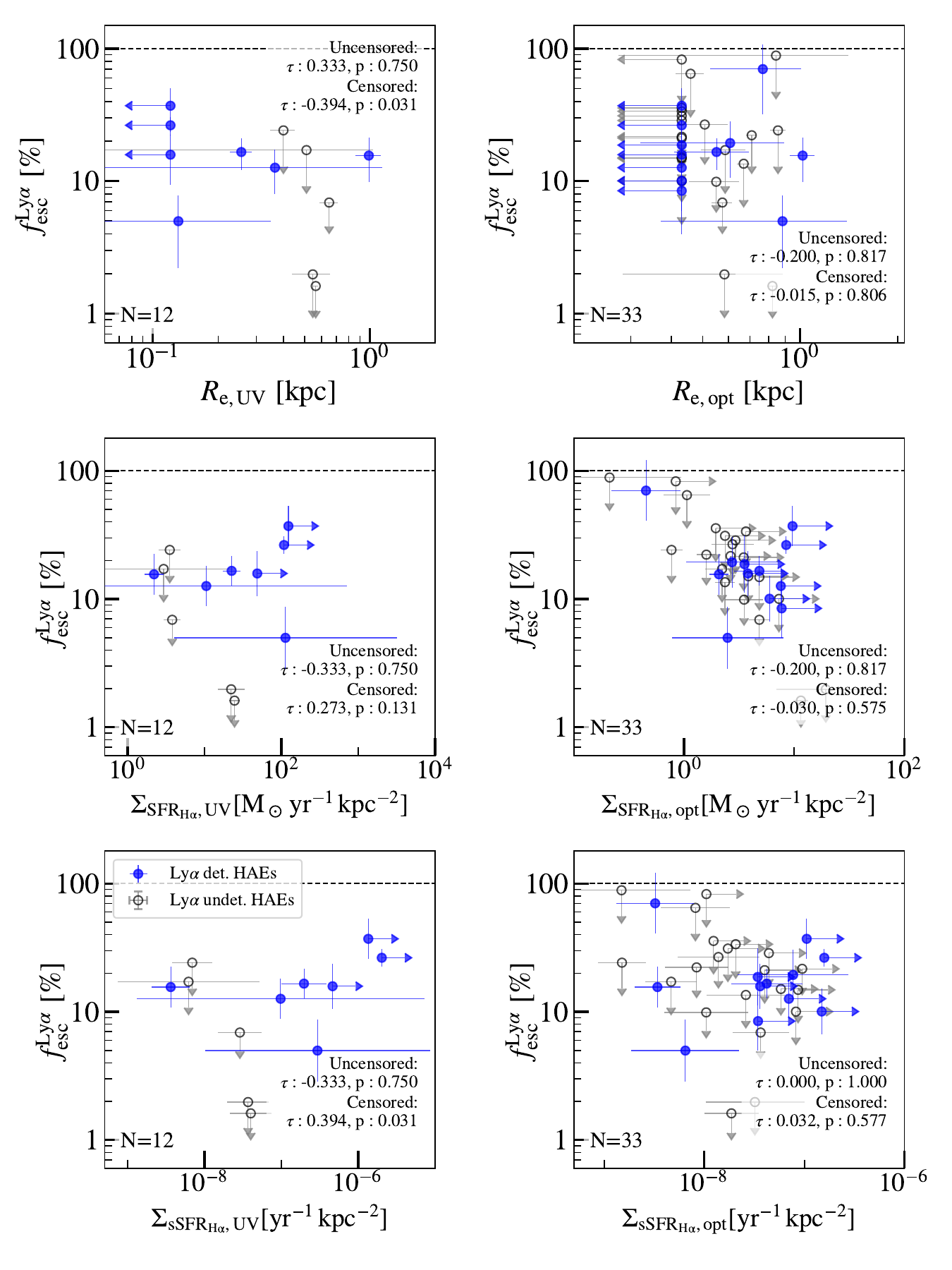}
    \caption{The relationships between \fesca\ and the rest-frame UV (left column) and optical (right column) effective radii (top row), $\Sigma_{\mathrm{SFR}}$ (middle row), and $\Sigma_{\mathrm{sSFR}}$ (bottom row) for HAEs at $z\simeq6.2$. The UV- and optical-size subsamples contain 12 and 33 HAEs, respectively. The surface densities are derived from H$\alpha$-based SFRs using the corresponding UV or optical effective radii. In each panel, filled blue circles represent Ly$\alpha$-detected HAEs (NB872 $>3\sigma$), while open grey circles indicate Ly$\alpha$-undetected HAEs. Downward arrows indicate $3\sigma$ upper limits on \fesca. Leftward arrows indicate resolution-limited upper limits on the effective radii, while rightward arrows indicate the corresponding lower limits on the surface densities. Error bars indicate $1\sigma$ uncertainties. The Kendall rank correlation coefficients $\tau$ and $p$-values are displayed for the `Uncensored'
    and `Censored' case, as defined in Figure~\ref{fig:f_esc_vs_physical_properties}.
}

    \label{fig:Sigma_SFR}
\end{figure}

We also investigate how \fesca\ depends on the rest-frame UV and rest-frame optical sizes (Figure~\ref{fig:Sigma_SFR}).
For sources whose fitted radii fall below the empirical measurement limits, we treat their sizes as upper limits in the censored Kendall tests.
We find a tentative negative trend between $R_{\mathrm{e,UV}}$ and \fesca, while no clear trend is found for $R_{\mathrm{e,opt}}$.
Notably, all three HAEs whose fitted UV radii fall below the empirical measurement limit are detected in NB872 and exhibit relatively high \fesca, qualitatively
supporting a connection between compact UV emission
and efficient \Lya\ escape.
However, given the small UV-size sample (12 HAEs, including five $f_{\mathrm{esc}}^{\mathrm{Ly\alpha}}$ upper limits and three resolution-limited UV sizes), as well as the large number of resolution-limited optical sizes (17/33), the UV-size trend remains tentative, and the apparent difference between the UV- and optical-size dependences should be interpreted with caution.

The rest-frame UV morphology is expected to trace the distribution of recent star formation that is visible along relatively low-attenuation sightlines, while the rest-frame optical morphology is more closely related to the underlying stellar continuum and hence to the $M_*$ distribution \citep{Ma+18,Yang+25_cosweb}.
The absence of a clear optical-size trend is also qualitatively consistent with the lack of a significant correlation between $M_*$ and \fesca\ noted above.
These results are consistent with the possibility that \Lya escape may be more closely linked to the distribution of the young star-forming components directly observable in the UV (i.e. the low-attenuation component traced by the UV continuum) rather than the overall galaxy structure, although larger samples are needed to test this interpretation.

A possible physical origin of such low-attenuation sightlines is stellar feedback associated with high SFR surface density, $\Sigma_{\mathrm{SFR}}$, and high specific SFR surface density, $\Sigma_{\mathrm{sSFR}}$.
The $\Sigma_{\mathrm{SFR}}$ quantifies how strongly star formation is concentrated per unit area within a galaxy and is commonly defined as
\begin{equation}
    \Sigma_{\mathrm{SFR}}=\frac{\mathrm{SFR_{H\alpha}}}{2\pi R_{\mathrm{e}}^2}.
\end{equation}
$\Sigma_{\mathrm{sSFR}}$ is defined as $\Sigma_{\mathrm{SFR}}$ divided by the $M_*$. 
Note that while $\Sigma_{\mathrm{sSFR}}$ is conventionally defined using stellar mass surface density ($\Sigma_*$), we adopt this definition to use $M_*$ as a proxy for the galaxy potential (e.g. \citealt{Reddy+22}).
Thus, it measures the intensity of star formation relative to the galaxy's gravitational potential.
We derive $\Sigma_{\mathrm{SFR}}$ and $\Sigma_{\mathrm{sSFR}}$ using $R_{\mathrm{e,UV}}$ and $R_{\mathrm{e,opt}}$ and examine their relation to the \fesca\ as shown in Figure~\ref{fig:Sigma_SFR}. 
The resolution-limited sizes correspond to lower limits on the derived surface densities, which are included in the censored Kendall tests.
We find tentative evidence for a positive correlation
between
$\Sigma_{\rm sSFR,UV}$ and \fesca, 
while no clear trends are found for the other surface-density quantities.
This trend is consistent with the picture in which \Lya escape is linked to concentrated star formation and its combination with a shallow potential.
Indeed, \citet{Reddy+22} analysed the variation of \Lya escape using interstellar absorption lines superimposed on the galaxy's own UV continuum, finding that \Lya escape is primarily governed by the covering fraction of optically thick \Hi.
They further showed that high $\Sigma_{\mathrm{sSFR}}$ can help reduce the covering fraction, thereby facilitating \Lya escape.
From a morphological perspective, LAEs are generally compact in the UV and exhibit high $\Sigma_{\mathrm{SFR}}$ \citep{Paulino-Afonso+18}, and arguments have been made linking compact morphologies and high $\Sigma_{\mathrm{SFR}}$ to \Lya escape via outflows and the formation of low-density channels \citep{Kim+25}.
Moreover, the connection between low covering fractions, which reflect a porous ISM,  and Ly$\alpha$/LyC escape is supported by detailed studies of nearby galaxies \citep{Gazagnes+20}.

We note that the $f_{\mathrm{esc}}^{\mathrm{Ly\alpha}}$ used in this study are effective
observed escape fractions, including not only radiative transfer through the ISM but also scattering and attenuation in the CGM and IGM. 
The scatter in Figures~\ref{fig:f_esc_vs_physical_properties} and
\ref{fig:Sigma_SFR}, including \Lya non-detections treated as upper limits in the censored Kendall tests, may therefore partly reflect line-of-sight variations in CGM/IGM transmission. 
At $z\simeq6.2$, such variations can arise from the inhomogeneous ionization topology and local ionized bubbles around galaxies (e.g. \citealt{Gronke+21,Meyer+25}). 
Future comparisons with simulations that combine galaxy-scale \Lya radiative transfer with
inhomogeneous CGM/IGM transmission (e.g. \citealt{zelda2}) will be needed to separate these effects from internal mechanisms regulating \Lya escape.

\section{Summary}\label{sec:summary}
We directly measure the \fesca\ of HAEs at $z\simeq6.2$ in the CEERS field, using a unique pair of NB filters: \textit{JWST}/NIRCam F470N (for H$\alpha$) and Subaru/HSC NB872 (for Ly$\alpha$). 
Through stacking analysis, we first measure the \fescmed\ and examine its dependence on the adopted lower \Ha\ luminosity limit. 
We then use individual \fesca\ measurements to investigate the physical conditions that regulate galaxy-to-galaxy variations in \Lya escape.
Our main results are summarized as follows:
\begin{enumerate}
    \item 
    A final sample of 84 HAEs at $z\simeq6.2$ is identified from F470N excess and photo-$z$ selections, after excluding three LRD candidates and one source with an uncertain HAE identification based on its SED.
    Among these HAEs, 56 have reliable NB872 photometry without severe
    contamination from nearby sources, 
    14 are detected in NB872 at $>3\sigma$.
    \item The HAEs are widely distributed on the star-forming MS and the SB region. In particular, HAEs with large \ewLya\ tend to show higher SFR than the MS, which suggests vigorous star formation activity. 
    Although many of the optical sizes are constrained only as upper limits, the measured optical sizes overlap with the size–M$_*$ relations for general SFGs.
    We find no clear dependence of galaxy size on \ewLya\ within the limitations of the current sample.
    \item Using stacking analysis, the \fescmed\ at $z\simeq6.2$ is measured as $0.090_{-0.032}^{+0.063}$, and $0.106_{-0.044}^{+0.066}$ when weighted to correct for detection incompleteness. 
    These measurements are consistent with previous results at $z\sim6$ based on LFs and spectroscopic stacking, and are much higher than analogous NB measurements at lower redshift, supporting an increase in the cosmic-averaged \fesca\ toward the EoR.
    \item No significant dependence of the stacked \fesca\ on the adopted lower \Ha\ luminosity limit is found within the luminosity range probed by our sample.
    If \Lya and LyC photons escape through similar low-density channels, this weak luminosity dependence may suggest that \fescc\ is not strongly dependent on H$\alpha$ luminosity over the range probed here.
    In that case, the ionizing photon budget 
    could receive a substantial contribution from 
    H$\alpha$-luminous galaxies 
    closer to the current observational limits. 
    \item Using individual measurements of \fesca, including upper limits for \Lya non-detections, the \fesca\ is found to show a positive correlation with \ewLya\ and a negative correlation with the UV slope $\beta$. 
    A tentative trend toward higher $f_{\mathrm{esc}}^{\mathrm{Ly\alpha}}$ at smaller rest-frame UV sizes is also observed, though the result is limited by the small sample and upper limits. 
    No significant correlations are found with rest-frame optical size, $M_*$, or SED-derived $E(B-V)$.
    These results suggest that the escape of \Lya photons may be more closely linked to the properties of the observable low-attenuation UV-emitting regions,
    \item 
    A tentative evidence for a positive correlation is found between UV-based $\Sigma_{\rm sSFR}$ and \fesca, while no clear correlations are found for the other surface-density quantities.
    Although based on a limited sample with some surface-density lower limits, this result is consistent with the possibility that outflows associated with concentrated star formation in compact regions reduce the covering fraction of the surrounding \Hi\ and create low-density channels acting as escape paths for \Lya photons.

\end{enumerate}

In this study, we apply the dual-NB stacking method to the EoR for the first time, demonstrating that it can still provide a direct measurement of the \fescmed\ for the \Ha-selected galaxies with low model dependence.
Future deeper or wider NB observations will be important for testing more robustly the \Ha\ luminosity dependence of the cosmic-averaged \fesca\ over a wider luminosity range.
They would also enable higher-S/N stacked \Lya and \Ha\ surface-brightness profiles, allowing us to compare the spatial distribution of resonantly scattered \Lya emission with that of recombination emission associated with recent star formation.
Such measurements would provide a direct way to separate the effects of galaxy-internal star formation and circumgalactic scattering, thereby offering deeper insights into the origin of \Lya haloes and the mechanisms of \Lya escape (e.g. \citealt{Leclercq+17,Kusakabe+22}).

The observed galaxy-to-galaxy variation in \fesca\ suggests that \Lya escape may be closely linked to the properties of compact, low-attenuation UV-emitting regions. 
However, larger samples will be needed to test these trends more robustly and to separate the effects of galaxy-internal processes from stochastic line-of-sight variations in IGM transmission.
Such samples will be essential for statistically averaging over variations in IGM attenuation and for identifying the galaxy-scale mechanisms that regulate \Lya escape. 
Moreover, the escape of \Lya photons is unlikely to be
controlled by a single physical quantity, but rather by a combination of UV slope $\beta$, size, $M_*$, $E(B-V)$, $\Sigma_{\mathrm{SFR}}$, $\Sigma_{\mathrm{sSFR}}$, gas geometry, and gas kinematics such as outflows.
Applying multivariate or interpretable machine-learning approaches, similar to those used by \citet{Yoshioka+25} to predict \Lya emission from multiple galaxy properties, to larger samples with direct \fesca\ measurements would provide a promising route to quantify the relative importance of the physical processes that govern \Lya escape during the EoR.

\section*{Acknowledgements}
We thank Naveen Reddy and Kazuhiro Shimasaku for constructive discussions.
SS is supported by the Japan Society for the Promotion of Science (JSPS) KAKENHI grant number JP26KJ0916.
NK is supported by the Japan Society for the Promotion of Science through Grant-in-Aid for Scientific Research 25H00663, 25K01038, 25K01044.
JA is supported by the Japan Society for the Promotion of Science (JSPS) KAKENHI grant number JP24KJ0858 and International Graduate Program for Excellence in Earth-Space Science (IGPEES), a World-leading Innovative Graduate Study (WINGS) Program, the University of Tokyo.
AI is supported by JSPS KAKENHI Grant Number 23H00131, 26H02069, 25K00020, 24H00002.
KI acknowledges support from the Independent Research Fund Denmark (DFF) under grant 3120-00043B. 
The Cosmic Dawn Center (DAWN) is funded by the Danish National Research Foundation under grant No. 140.
SK is supported by the JSPS KAKENHI grant number 24KJ0058 and 24K17101.
MK is supported by the JSPS KAKENHI grant number JP25K01032.
KN acknowledges support from JSPS KAKENHI grant 20H00180, 24H00002, 24H00241, JP25K01032, and the JSPS International Leading Research (ILR) project, JP22K21349.
KN also acknowledges support from the Kavli IPMU, the World Premier Research Center Initiative (WPI), UTIAS, and the University of Tokyo.
MO is supported by the Japan Society for the Promotion of Science (JSPS) KAKENHI Grant Number 24K22894.
RS is also supported by the JSPS KAKENHI grant number JP25K01044. 

The Hyper Suprime-Cam (HSC) collaboration includes the astronomical communities of Japan and Taiwan, and Princeton University. The HSC instrumentation and software were developed by the National Astronomical Observatory of Japan (NAOJ), the Kavli Institute for the Physics and Mathematics of the Universe (Kavli IPMU), the University of Tokyo, the High Energy Accelerator Research Organization (KEK), the Academia Sinica Institute for Astronomy and Astrophysics in Taiwan (ASIAA), and Princeton University. Funding was contributed by the FIRST program from the Japanese Cabinet Office, the Ministry of Education, Culture, Sports, Science and Technology (MEXT), the Japan Society for the Promotion of Science (JSPS), Japan Science and Technology Agency (JST), the Toray Science Foundation, NAOJ, Kavli IPMU, KEK, ASIAA, and Princeton University. 

This paper makes use of software developed for Vera C. Rubin Observatory. We thank the Rubin Observatory for making their code available as free software at http://pipelines.lsst.io/.

This paper is based on data collected at the Subaru Telescope and retrieved from the HSC data archive system, which is operated by the Subaru Telescope and Astronomy Data Center (ADC) at NAOJ. Data analysis was in part carried out with the cooperation of Center for Computational Astrophysics (CfCA), NAOJ. We are honored and grateful for the opportunity of observing the Universe from Maunakea, which has the cultural, historical and natural significance in Hawaii. 

The Pan-STARRS1 Surveys (PS1) and the PS1 public science archive have been made possible through contributions by the Institute for Astronomy, the University of Hawaii, the Pan-STARRS Project Office, the Max Planck Society and its participating institutes, the Max Planck Institute for Astronomy, Heidelberg, and the Max Planck Institute for Extraterrestrial Physics, Garching, The Johns Hopkins University, Durham University, the University of Edinburgh, the Queen’s University Belfast, the Harvard-Smithsonian Center for Astrophysics, the Las Cumbres Observatory Global Telescope Network Incorporated, the National Central University of Taiwan, the Space Telescope Science Institute, the National Aeronautics and Space Administration under grant No. NNX08AR22G issued through the Planetary Science Division of the NASA Science Mission Directorate, the National Science Foundation grant No. AST-1238877, the University of Maryland, Eotvos Lorand University (ELTE), the Los Alamos National Laboratory, and the Gordon and Betty Moore Foundation.

This work is based in part on observations made with the NASA/ESA/CSA James Webb Space Telescope. The data were obtained from the Mikulski Archive for Space Telescopes at the Space Telescope Science Institute, which is operated by the Association of Universities for Research in Astronomy, Inc., under NASA contract NAS 5-03127 for JWST. These observations are associated with program \#2234.

We acknowledge the use of GitHub Copilot, ChatGPT (OpenAI), and Google Gemini for English-language editing, wording refinement, and translation of parts of the manuscript. 
The tools were used only to improve clarity and readability.
We reviewed all AI-assisted text and takes full responsibility for the final manuscript.

\section*{Data Availability}
All \textit{HST}/\textit{JWST} data, including the F470N data, are available from the Mikulski Archive for Space Telescopes (MAST) at \url{https://archive.stsci.edu/}. The NB872 data and the associated catalogues are available upon reasonable request.



\bibliographystyle{mnras}
\bibliography{ref} 




\appendix

\section{Additional correlation between galaxy properties and the Ly$\alpha$ escape fraction}
\label{app:additional_correlations}
\begin{figure*}
    \centering
    \includegraphics[width=0.7\textwidth]{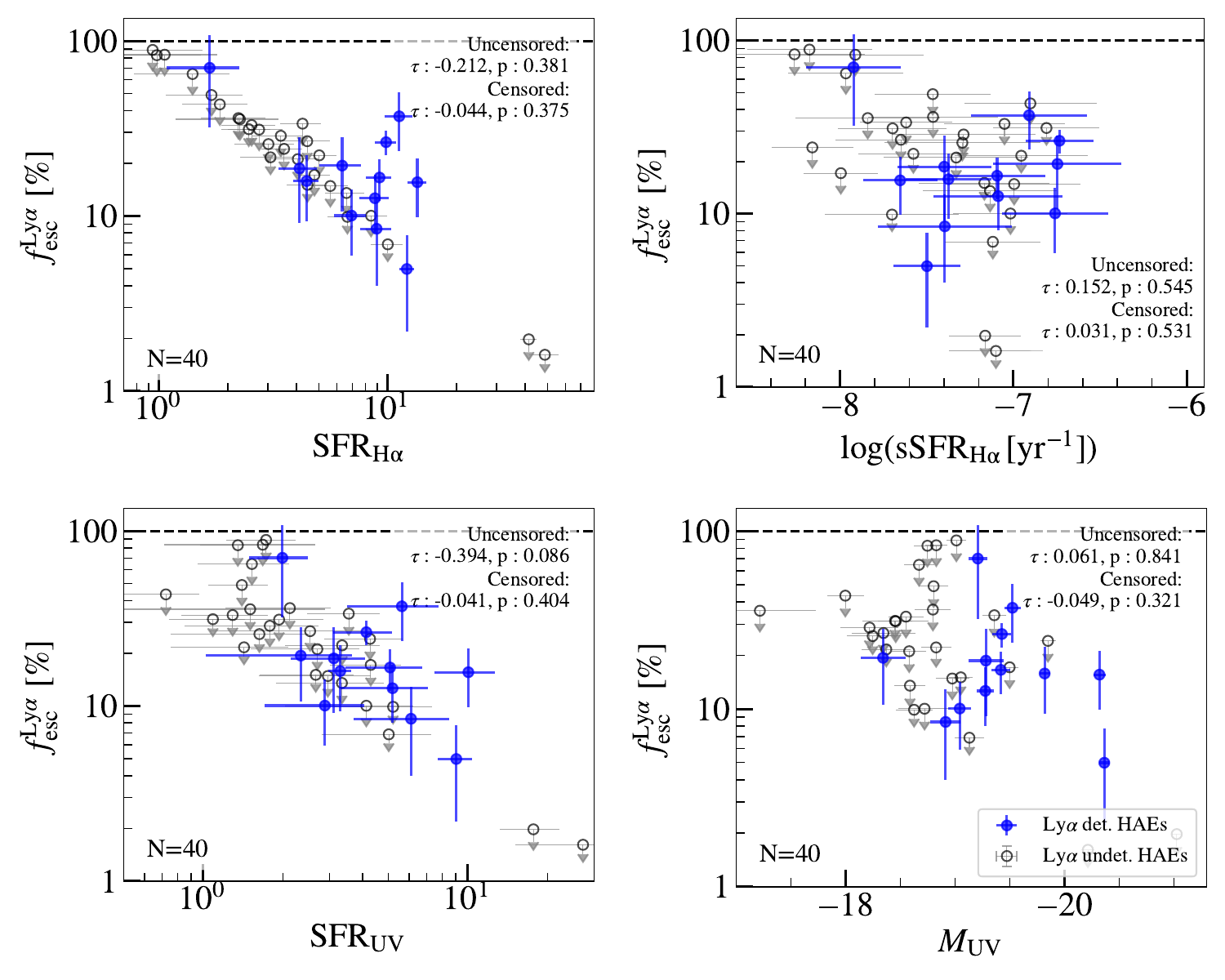}
    \caption{Same as Fig. 12, but showing the relationships between $f_{\mathrm{esc}}^{\mathrm{Ly\alpha}}$ and galaxy properties for 40 HAEs at $z\simeq6.2$: $\mathrm{SFR_{H\alpha}}$, $\mathrm{sSFR_{H\alpha}}$, $\mathrm{SFR_{UV}}$, and $M_{\rm UV}$. 
    }

    \label{fig:f_esc_vs_add}
\end{figure*}
We additionally investigate possible relationships between $f_{\mathrm{esc}}^{\mathrm{Ly\alpha}}$ and $\mathrm{SFR}_{\mathrm{H\alpha}}$, $\mathrm{sSFR}_{\mathrm{H\alpha}}$, $\mathrm{SFR}_{\mathrm{UV}}$, and $M_{\rm UV}$.
The $\mathrm{SFR}_{\mathrm{H\alpha}}$ is calculated as described in Section~\ref{sec:f_esc_Lya_calculation}, and $\mathrm{sSFR}_{\mathrm{H\alpha}}$ is defined as $\mathrm{SFR}_{\mathrm{H\alpha}}/M_*$.
We derive $M_{\mathrm{UV}}$, the rest-frame UV absolute magnitude at 1500\,\AA, from the best-fitting SED model, in the same manner as the 1216\,\AA\, continuum used for the \Lya flux measurement.
To calculate $\mathrm{SFR}_{\mathrm{UV}}$, we first correct the 1500\,\AA\, luminosity for dust attenuation using the SED-derived $E(B-V)$ and the Calzetti attenuation law \citep{Calzetti+00}, as adopted for the \Ha\ luminosity.
We then convert the dust-corrected UV luminosity into $\mathrm{SFR}_{\mathrm{UV}}$ following \citet{Kennicutt+98}, adapted for a \citet{Chabrier+03} IMF, as in \citet{Shimizu+25}.
Figure~\ref{fig:f_esc_vs_add} presents the relationships between these four quantities and \fesca.
No statistically significant correlations are found for any of these quantities.


\bsp	
\label{lastpage}
\end{document}